\documentclass[aps,prd,groupedaddress,showpacs,nofootinbib,amssymb,balancelastpage,preprintnumbers,floatfix]{revtex4-2}

\usepackage[dvipdfmx]{graphicx}
\usepackage{graphicx,bm,color}

\usepackage{amsmath}
\usepackage{amssymb}
\usepackage{amsfonts}
\usepackage{cases}
\usepackage{cancel}
\usepackage{hyperref}
\usepackage[normalem]{ulem}
\usepackage{subfigure}
\usepackage{here}
\usepackage{color}
\usepackage{comment}
\usepackage{wasysym}

\usepackage{tikz}
\usetikzlibrary{matrix}

\allowdisplaybreaks[3]

\begin{document}
	

\title{ 
Significance of soft-scale breaking on primordial black hole production 
in Coleman-Weinberg type supercooling-phase transition 
}

\author{He-Xu Zhang}\thanks{{\tt zhanghexu@ucas.ac.cn}}
\affiliation{School of Nuclear Science and Technology, University of Chinese Academy of Sciences, Beijing 100049, China}

\author{Katsuya Hashino}\thanks{{\tt hashino@fukushima-nct.ac.jp}}
\affiliation{National Institute of Technology, Fukushima College, 
Fukushima 970–8034, Japan}

\author{Hiroyuki Ishida}\thanks{{\tt ishidah@pu-toyama.ac.jp}}
\affiliation{Center for Liberal Arts and Sciences, Toyama Prefectural University, Toyama 939-0398, Japan}
\author{Shinya Matsuzaki}\thanks{{\tt synya@jlu.edu.cn}}
\affiliation{Center for Theoretical Physics and College of Physics, Jilin University, Changchun, 130012,	China}%

\begin{abstract}
Ultra-supercooling phase transitions can generate large overdensities in the Universe, potentially leading to the formation of primordial black holes (PBHs), which can also be a dark matter candidate. In this work, we focus on the supercooling phase transition for the scale symmetry breaking based on the effective potential of the Coleman-Weinberg (CW) type. 
We investigate the effect on the PBH production in the presence of 
an additional mass term for the CW scalar field, what we call a soft-scale breaking term, which serves as the extra explicit-scale breaking term other than the quantum scale anomaly induced by the CW mechanism. 
We demonstrate that even a small size of the soft-scale breaking term  
can significantly affect the PBH production depending on its sign: 
a positive term slows down the phase transition, thereby enhancing the PBH abundance and improving the model’s ability to account for dark matter; in contrast, a negative term suppresses the PBH formation. 
The inclusion of such soft-scale breaking terms broadens the viable parameter space and increases the flexibility of the framework. We further illustrate our results through two ultraviolet-complete realizations: i) a many-flavor QCD-inspired model as a reference model which can dynamically induce a positive-soft scale breaking; ii) a Higgs portal model with a $B-L$ scalar as the benchmark for the case where a negative-soft scale breaking is induced. Our study would provide a new testable link between PBH dark matter and gravitational wave signatures in the CW-type scenario. 
	
\end{abstract}

\maketitle

\section{Introduction}

Primordial black holes (PBHs), hypothesized to form in the early Universe from large density perturbations~\cite{Zeldovich:1967lct,Hawking:1971ei}, have recently garnered renewed interest as viable dark matter candidates. 
The persistent null results in experimental searches for weakly interacting massive particles and axions~\cite{DiValentino:2014zna}, namely the particle dark matter candidates, have further strengthened the case for PBHs as a non-particle alternative~\cite{Carr:2020gox}.
In particular, PBHs in the asteroid-mass range $\sim (10^{-16} - 10^{-10}) M_\odot$, remain unconstrained by evaporation and microlensing bounds~\cite{Niikura:2017zjd,Katz:2018zrn,Smyth:2019whb,Montero-Camacho:2019jte}. 
These PBHs could potentially constitute all of the dark matter, making this mass window a region of significant interest from a phenomenological perspective. 

One particularly intriguing mechanism for the PBH formation is \textit{delayed vacuum transitions}, which can occur during supercooling phase transitions in the early Universe~\cite{Kodama:1982sf,Liu:2021svg}. In this mechanism, bubble nucleation, a stochastic process, causes variations in timing across different Hubble patches. In patches where nucleation is delayed, the vacuum energy remains nearly constant, while surrounding patches where nucleation has occurred convert this energy into radiation, which dilutes as the Universe expands. The resulting energy density contrast can create significant local overdensities, and if these exceed a critical threshold, the enclosed mass can gravitationally collapse to form PBHs.
This mechanism strongly motivates the study of the supercooling phase transition, which naturally yields small values of the inverse duration parameter $\beta$ and the prolonged period of vacuum energy domination before bubble percolation, during which large density contrasts can develop. Moreover, the extended transition duration increases the likelihood of delayed Hubble patches, enhancing the probability of PBH formation.

The most widely studied supercooling models can be categorized in a view of the physical origin of the requisite potential barrier. 
Given a scalar field that is profiled by the target potential, 
a class of \textit{tree-level driven}
models features a barrier arising from a competition between terms in the effective potential that are already present at tree level, often involving cubic interactions for the scalar field. A typical approach to engineer such models is to tune the tree-level couplings by hand, so as to generate a sufficiently high and wide barrier that suppresses the tunneling rate. In such scenarios, the bounce action $S_3(T)/T$ can exhibit a characteristic ``U-shaped'' behavior as a function of temperature, indicating a prolonged period of supercooling. The PBH production in this context has been investigated by Xie et al.~\cite{Kanemura:2024pae}, as well as in similar models discussed in~\cite{Goncalves:2024vkj}.

In contrast, a \textit{radiatively driven} potential barrier is realized via the Coleman-Weinberg (CW) mechanism~\cite{Coleman:1973jx} in classically scale-invariant (CSI) theories. In this framework, all dimensionful parameters are absent at tree level, but quantum loop corrections induce the scale anomaly and generate the CW potential leading to the spontaneous scale-symmetry breaking. 
Since the CW potential merely includes a logarithmic field dependence, it exponentially suppresses the tunneling rate at low temperatures. This intrinsic feature naturally induces strong enough supercooling without tuning, provided the couplings are not too large, making these models particularly relevant for the PBH formation.

CSI extensions of the Standard Model (SM), 
which encode the CW mechanism, 
are attractive since they not only provide a potential resolution to the hierarchy problem in particle physics via the dimensional transmutation, but also have been successfully applied to cosmological contexts, including the so-called CW-type small-field inflation~\cite{Zhang:2023acu,Tambalo:2016eqr} and the generation of a stochastic gravitational wave (GW)  background~\cite{Jaeckel:2016jlh,Marzola:2017jzl,Iso:2017uuu,Ghorbani:2017lyk,Baldes:2018emh,Prokopec:2018tnq,Brdar:2018num,Marzo:2018nov,DelleRose:2019pgi,Ghoshal:2020vud,Levi:2022bzt,Kierkla:2022odc,Salvio:2023qgb,Zhang:2024vpp}, hence making them both theoretically and cosmologically well-motivated.

Recently, CSI theories have received considerable attention in a view of the PBH formation via the mechanism of delayed vacuum transitions~\cite{Gouttenoire:2023naa,Gouttenoire:2023pxh,Baldes:2023rqv,Conaci:2024tlc,Arteaga:2024vde,Salvio:2023ynn,Salvio:2023blb,Banerjee:2024cwv,Kierkla:2025vwp}.
According to semi-analytical estimates~\cite{Gouttenoire:2023naa}, in the ultra-supercooling regime ($\alpha \gtrsim 10^2$), explaining the observable dark matter via PBHs requires the parameter $\beta/H$ to lie within the range of $\sim 6.4 - 6.9$. 
If $\beta/H$ falls below this range, PBHs would be overproduced and rapidly dominate the energy density of the universe. Conversely, if $\beta/H$ exceeds this range, corresponding to a rapid phase transition, the resulting PBH abundance would be insufficient to account for dark matter. 

However, this result is subject to the assumption of the exponential nucleation approximation $\Gamma \sim\mathrm{e}^{-\beta t}$, which may not hold in all scenarios. 
For instance, if $\beta/H$ approaches zero or becomes negative,  
the exponential nucleation approximation breaks down, and the improved nucleation treatment must be adopted. 
Consequently, the semi-analytical fitting formulas used in previous studies to estimate PBH abundances are no longer valid. 
Actually, a soft-scale breaking-mass term ($ V_{m_0} \sim \pm m_0^2 \phi^2$) for the CW scalar field ($\phi$) can realize this situation, 
in which case the formation of PBHs warrants a careful re-examination.

In this paper, we investigate the effect on the PBH production in the presence of 
an additional mass term for the CW scalar field, $V_{m_0} \sim \pm m_0^2 \phi^2$, 
what we call a soft-scale breaking term, which serves as the extra explicit-scale breaking term other than the quantum 
scale anomaly induced by the CW mechanism. 
We show that even a small size of the soft-scale breaking term  
can significantly affect the PBH production depending on its sign: a positive term (with $V_{m_0} \sim + m_0^2\phi^2$) slows down the phase transition, thereby enhancing the PBH abundance and improving the ability of the model to account for dark matter; in contrast, a negative term (with $V_{m_0} \sim - m_0^2 \phi^2$) suppresses the PBH formation.

As benchmark models, we introduce two scenarios: i) a many-flavor QCD-inspired model as a reference model which can dynamically induce a positive-soft scale breaking ($V_{m_0} \sim + m_0^2 \phi^2$); ii) a Higgs portal model with a $B-L$ scalar as the benchmark for the case where a negative-soft scale breaking ($V_{m_0} \sim - m_0^2\phi^2$) is induced. 
Our present study would thus provide a new testable link between PBH dark matter and gravitational wave signatures in the CW-type scenario.

This paper is organized as follows: in Sec.~\ref{sec:model}, we discuss generic properties of CSI models of CW type by taking a reference model. 
Section~\ref{sec:PT} focuses on the analytical derivation of key formulas describing the dynamics of the ultra-supercooling phase transition. 
In Sec.~\ref{sec:PBHs}, we delve into the formation of PBHs and its related phase transition dynamics. 
Section~\ref{sec:Significance} highlights the significance of soft-scale breaking mass terms with plus or minus sign, where we also analyze its impact on the PBH abundance and related parameters. 
In Secs.~\ref{sec:walking-model} and~\ref{sec:B-L-model}, we present two reference models that realize positive and negative soft-scale breaking mass terms, respectively. 
The summary of our present study is given in Sec.~\ref{sec:summary}.

\section{The target type of the model}\label{sec:model}

To monitor generic properties of CSI models of CW type, 
we consider the CSI extension of the SM along with the 
$U(1)_D$ dark gauge field $V_\mu$ and 
the Higgs portal partner $\Phi$ charged under the $U(1)_D$. 
The Lagrangian up to mass dimension four is given as 
\begin{equation}
    \mathcal{L} = \mathcal{L}_{\overline{\mathrm{SM}}} -\frac{1}{4}F_{\mu\nu}^2 + \left|D_\mu\Phi\right|^2 - V_{\rm tree}(H,\Phi)\,,
\end{equation}
with the scalar potential part, 
\begin{equation}
    V_{\rm tree}(H,\Phi) = \lambda_h |H|^4 - \lambda_{h\phi} |H|^2|\Phi|^2 + \lambda_\phi |\Phi|^4\,.
\end{equation}
Here $F_{\mu\nu} = \partial_\mu V_\nu - \partial_\nu V_\mu$,  
and the covariant derivative acting on the $U(1)_D$ scalar $\Phi$ is defined as 
$D_\mu \Phi = (\partial_\mu + i g_D V_\mu) \Phi$ with 
$g_D$ being the gauge coupling constant and the $U(1)_D$ charge for $\Phi$ set to +1, for simplicity. 

Assuming $g_D^2 \gg \lambda_\phi,\,\lambda_{h\phi}$, the potential develops a sufficiently flat direction at tree level and the nontrivial vacuum at one-loop level via the CW mechanism \cite{Coleman:1973jx}. 
This assumption allows one to neglect 
the $\Phi$-scalar self- ($\lambda_\phi$) and Higgs portal ($\lambda_{h \phi}$) interactions at one-loop level, in which the latter ensures decoupling of the SM sector from the $U(1)_D$ sector at this point. 
By working in the Landau gauge for $U(1)_D$, 
the one-loop effective potential for $\phi = \sqrt{2} |\Phi|$ 
in the $\overline{\rm MS}$ renormalization scheme is then evaluated to be 
\begin{equation}
	\label{zeropotential}
	V_{\rm eff}(\phi)= \frac{N_b}{64\pi^2}m_V^4(\phi)\left(\log\frac{m_V^2(\phi)}{\mu_R^2}-\frac{3}{2}\right),\quad m_V(\phi) = g_D \phi \,,
\end{equation} 
where $N_b=3$ denotes the number of degrees of freedom (d.o.f.) of the $U(1)_D$ gauge boson, and $\mu_R$ is the renormalization scale. 

The stationary condition leads to the dimensional transmutation for $g_D$:
\begin{equation}
	\left.\frac{\partial V_{\rm eff}(\phi)}{\partial \phi}\right|_{\phi=v_\phi}=0 \quad \Rightarrow \quad \mu_R = g_D v_\phi/\mathrm{e}^{1/2}\,.
\end{equation}
Plugging this condition into Eq.~(\ref{zeropotential}), one gets 
\begin{equation}
\label{Veffwithvacuum}
	V_{\rm eff} (\phi)= \frac{N_bg_D^4\phi^4}{32\pi^2}\left(\log\frac{\phi}{v_\phi}-\frac{1}{4}\right) + V_0\,,
\end{equation}
where $V_0$ is the vacuum energy normalized as $V_{\rm eff}(\phi=v_\phi)=0$ and $V_0 = N_bg_D^4v_\phi^4/(128\pi^2)$.

To investigate the thermal phase transition, we incorporate the one-loop thermal corrections in the $U(1)_D$ sector into the effective potential to get 
\begin{equation}
	\label{eq:thermal-potential}
	V_{\rm eff} (\phi,T)=  \frac{N_bg_D^4\phi^4}{32\pi^2}\left(\log\frac{\phi}{v_\phi}-\frac{1}{4}\right) + \frac{N_b T^4}{2\pi^2} J_B\left(\frac{g_D^2\phi^2}{T^2}\right) + V_0\,,
\end{equation}
where $J_B$ is the bosonic thermal function given as 
\begin{align} 
J_B(y^2) = \int_{0}^{\infty}t^2 \ln\left(1-\mathrm{e}^{-\sqrt{t^2+y^2}}\right)\mathrm{d}t\,. 
\end{align} 
To control the high temperature corrections properly, it is necessary to also incorporate an additional thermal correction to the effective potential arising from the resummation of daisy diagrams \cite{Carrington:1991hz,Arnold:1992rz}. However, as was demonstrated in \cite{Carrington:1991hz,Kawana:2022fum}, 
in the ultra-supercooling scenario as in the present model case, 
the difference between the thermal effective potentials with and without daisy resummation becomes insignificant, as the phase transition typically occurs well below the critical temperature. 
Therefore, we simply neglect the daisy diagrams contributions at this point, which will  be recovered in performing the numerical analysis later on.

The framework presented here can straightforwardly be extended to other theoretical models, such as the conformal $SU(2)_D$ extension of the SM~\cite{Hambye:2018qjv,Frandsen:2022klh,Arteaga:2024vde}. In this case, the generalization involves replacing the coupling constant $g_D$ with $g_D/2$, and the setting the number of bosonic d.o.f. $N_b=9$ in Eq.~(\ref{eq:thermal-potential}). Similarly, the framework can be adapted to the Gildener-Weinberg model~\cite{Gildener:1976ih} by applying the substitutions for $g_D = \sqrt{\lambda}$, where $\lambda$ is the interaction strength between $N_b$ scalar fields.
In that sense, what will follow in the later section can be interpreted as essential features on the general phase transition dynamics of the CW type.

\section{Ultra-supercooling phase transition of CW type with or without soft-scale breaking mass}\label{sec:PT}

In this section, we discuss the characteristics of cosmological phase transitions arising from the CW-type framework, in the cases, respectively, with and without a soft-scale breaking mass term. 

\subsection{The case without soft-scale breaking mass} 

In the expanding universe the first order phase transition proceeds by the bubble nucleation.  
The nucleation rate per unit volume/time of the bubble, $\Gamma(T)$, 
can be computed as 
\begin{equation}
	\label{nucleation-rate}
	\Gamma(T) \simeq T^4\left(\frac{S_3(T)}{2\pi T}\right)^{3/2}\exp\left(-\frac{S_3(T)}{T}\right),
\end{equation}
where $S_3(T)$ is the $\mathcal{O}(3)$ symmetric bounce action at $T$: 
\begin{equation}
	S_3(T) = 4\pi\int_0^\infty d r\, r^2\left(\frac{1}{2}\left(\frac{\mathrm{d}\bar{\phi}}{\mathrm{d}r}\right)^2+V_{\rm eff}(\bar{\phi},T)\right).
\end{equation}
The normalizable bubble profile $\bar{\phi}(r)$ can be obtained by numerically solving the equation of motion,  
\begin{equation}
	\frac{\mathrm{d^2}\bar{\phi}}{\mathrm{d}r^2}+\frac{2}{r}\frac{\mathrm{d}\bar{\phi}}{\mathrm{d}r}=\frac{\mathrm{d}V_{\rm eff}(\bar{\phi},T)}{\mathrm{d}\bar{\phi}}
	\,, 
\end{equation}
with the boundary condition 
\begin{align}
	\left.\frac{2}{r}\frac{\mathrm{d}\bar{\phi}(r)}{\mathrm{d}r}\right|_{r=0}=0,\quad \left.\bar{\phi}(r)\right|_{r \to \infty}=0\,. 
\end{align}

To derive a more handy formula for the nucleation rate, we may work 
on high-$T$ expansion, only keeping the thermal mass term $\propto T^2$. 
This still seems to work when $\phi$ goes away from the scale-symmetric (false) vacuum at $\phi=0$, as long as $\phi < T/g_D$, i.e. the $U(1)_D$ gauge-thermal loop is still active. 
This is applicable to the case where we are particularly concerned about the potential barrier generation in the CW type, which is seen at the place distant from the origin due to the approximate scale invariance (flatness) around there. 
Thus the effective potential in Eq.~(\ref{eq:thermal-potential}) can be simply evaluated as 
\begin{equation}
    V_{\rm eff}(\phi, T) \Bigg|_{\phi \lesssim \frac{T}{g_D}} \simeq  \frac{N_b g_D^2 T^2 \phi^2}{24} + \frac{N_bg_D^4\phi^4}{32\pi^2}\left(\log\frac{\phi}{v_\phi}-\frac{1}{4}\right)\,,
\end{equation}
where the field-independent terms have been omitted. 
In the supercooling case with $T\ll v_\phi$, 
the dominant contribution to the bounce action comes from the field values near the barrier, where $\phi \sim T/g_D$. 
Consequently, the logarithm term above can be expressed as
\begin{equation}
	\ln\frac{\phi}{v_\phi} \Bigg|_{g_D \phi \sim T } = \left( \ln\frac{g_D\phi}{T} + \ln\frac{T}{g_D v_\phi} \right)
    \Bigg|_{g_D \phi \sim T } \sim \ln\frac{T}{g_D v_\phi}\equiv \ln\frac{T}{M}\,,
\end{equation}
where $M$ stands for the $U(1)_D$ gauge boson mass scale, which is also 
the typical mass scale at the true vacuum. 
Thus, the effective potential can be recast in a concise form:  
\begin{equation}
	 V_{\rm eff}(\phi, T)\Bigg|_{\phi \sim \frac{T}{g_D}} \simeq \frac{m^2(T)}{2} \phi^2 - \frac{\lambda_4(T)}{4}\phi^4\,,
\label{concise-V}
\end{equation}
with
\begin{equation}
	m^2(T) = \frac{N_b g_D^2 T^2}{12}\,, \qquad  \lambda_4(T) = \frac{N_b}{8\pi^2}g_D^4\ln\frac{M}{T}\,.
\end{equation}

For $m^2(T)>0$ and $\lambda_4(T)>0$, the thick-wall formula provides a reliable approximation for the bounce action~\cite{Brezin:1992sq,Witten:1980ez}. 
When the approximation is applied to the effective potential in Eq.~(\ref{concise-V}), the bounce action reads 
\begin{equation}
	S_3(T)\simeq 18.897 \frac{m(T)}{\lambda_4(T)}\,, \qquad 
    \textrm{or equivalently} \qquad \frac{S_3(T)}{T} \simeq  \frac{430.72}{\sqrt{N_b}g_D^3}\frac{1}{\ln\frac{M}{T}}\,.
\label{S3/Tn-handy}
\end{equation}
Thus, in the approximately scale-invariant framework, the bounce action decreases logarithmically with temperature. As a result, the tunneling rate $\Gamma(T) \propto \exp(-S_3/T)$ remains exponentially suppressed over a wide range of temperatures, preventing bubble nucleation from the supercooling phase transition.

The nucleation temperature $T_n$ is defined at which one bubble is nucleated in one casual Hubble 
\begin{equation}
	N(T_n)=\int_{t_c}^{t_n}\,\mathrm{d}t\frac{\Gamma(t)}{H(t)^3}=\int_{T_n}^{T_c}\, \frac{\mathrm{d}T}{T}\frac{\Gamma(T)}{H(T)^4}=1\,,
\end{equation}
where $T_c$ is the critical temperature at which the false ($\phi=0$) and true ($\phi=v_\phi$) vacua are degenerate.
For the exponential nucleation rate, this degeneracy happens at the moment 
when the bubble nucleation rate for the first time catches up with the Hubble expansion rate:   
\begin{equation}
	\frac{\Gamma(T_n)}{H(T_n)^4} \simeq 1\,, 
    \qquad {\rm i.e., when} \qquad \frac{S_3(T_n)}{T_n}-\frac{3}{2}\ln\left(\frac{S_3(T_n)}{2\pi T_n}\right) 
		\simeq 4\ln\frac{T_n}{H(T_n)}\,.
\end{equation}

In the supercooling case, the Hubble parameter $H(T)$ can be well approximated by 
the vacuum energy part $H_0^2=V_0/(3M_{\rm pl}^2)$, where $M_{\rm pl}$ is the reduced Planck mass $\simeq 2.4\times 10^{18} \text{ GeV}$. Thus, the nucleation temperature is determined by 
\begin{equation}
\label{TnforCW}
	T_n \simeq \sqrt{M H_0} \exp\left(\frac{1}{2}\sqrt{\ln^2\left(\frac{M}{H_0}\right)-\frac{430.72}{\sqrt{N_b}g_D^3}}\right).
\end{equation}
One can thus find a lower bound on $T_n$ at the minimum coupling $g_D^{\rm min}$ 
\begin{equation}
	T_n^{\rm min} \simeq \sqrt{M H_0} \sim g_D^{\rm min}v_\phi\left(\frac{g_D^{\rm min}v_\phi}{M_{\rm pl}}\right)^{1/2}
    = M^{\rm min} \left( \frac{M^{\rm min}}{M_{\rm pl}} \right)^{1/2}
    \,.
\end{equation}

To characterize the strength of the phase transition, we introduce the trace anomaly difference between the two vacua, defined as
\begin{equation}
    \Delta\theta(T) = \Delta V(T) - \frac{1}{4}T\frac{\partial \Delta V(T)}{\partial T}\,,
\end{equation}
where $\Delta V(T) = V(\phi=0,T) - V(\phi=v_\phi,T)$. 
The phase-transition strength parameter $\alpha$ is then given as 
\begin{equation}
    \alpha = \frac{\Delta\theta(T_n)}{\rho_r(T_n)} \simeq \frac{V_0}{(\pi^2/30) g_\star(T_n) T_n^4} = \frac{15\, N_b  M^4  
    }{64\pi^4\, g_\star(T_n) T_n^4}\,,
\end{equation}
which measures the relative energy released during the phase transition with respect to the radiation energy in the thermal plasma. 
Here, $g_\star(T)$ denotes the effective number of relativistic d.o.f. in the symmetric phase at temperature $T$. 
$g_\star(T_n)$ includes the massless SM contribution and may or may not also have $U(1)_D$ gauge-Higgs sector contributions, depending on thermal history of the $U(1)_D$ sector, 
which we will not explicitize at this point.

As was pointed out in~\cite{Zhang:2024vpp}, 
in the ultra-supercooling of the CW-type 
the nucleation temperature simply scales with $v_\phi$:
$T_n \propto v_\phi$, which remains valid when $v_\phi\ll M_{\rm pl}$.
This implies that $\alpha$ does not almost have the dependence of $v_\phi$. This trend can also be seen from Eq.~(\ref{TnforCW}) by considering $M/H_0$ to be large enough.

Another key parameter, $\beta$, characterizing the duration of the phase transition, is defined as $\beta \equiv \mathrm{d} \ln \Gamma / \mathrm{d}t$, which, when normalized by the Hubble parameter at $T=T_n$ as $\beta_H = \beta/H$, is evaluated as   
\begin{equation}
	\beta_H \equiv \frac{\beta}{H(T_n)} \simeq  -4+\left.T \frac{\partial(S_3/T)}{\partial T}\right|_{T=T_n} = -4 + \frac{1}{\ln(M/T_n)}\frac{S_3}{T_n} =  -4+\frac{\sqrt{N_b}g_D^3}{430.72}\left(\frac{S_3}{T_n}\right)^2\,. 
\label{betaH}
\end{equation}
Here we have used the adiabatic time–temperature relation $\mathrm{d}t = -\mathrm{d}T / (T H(T))$ and Eq.~(\ref{S3/Tn-handy}) in reaching the last equality.
The thermal prefactor $T^4$ in Eq.~(\ref{nucleation-rate}) accounts for the ($-4$) term in $\beta_H$, which becomes important when the phase transition is slow, i.e., when $\beta_H$ is small. In contrast, the contribution from a factor of $(S_3/2\pi T)^{3/2}$ in Eq.~(\ref{nucleation-rate}) is generally subleading, typically of order $\mathcal{O}(10^{-1})$ compared to the exponential term $\exp(-S_3/T)$, because $S_3/T \sim {\cal O}(10^2)$ when the nucleation takes place. 
This term contribution has thus been neglected in the estimation of $\beta_H$ for simplicity. 

It should also be noted that due to the ultra-supercooling, the nucleation temperature $T_n$ becomes much smaller than the mass scale $M$, resulting in a large suppression factor against $S_3/T_n$ of ${\cal O}(10^2)$ in Eq.~(\ref{betaH}). 
This feature drives $\beta_H$ to approach values of order $\mathcal{O}(1)$, thereby enables the production of a sufficient abundance of PBHs to account for dark matter.

\subsection{The case with soft-scale breaking mass}

Consider a small size of the soft-scale-breaking mass term, 
\begin{equation}
	V_{m_0} = \pm \frac{m_0^2}{2}\phi^2\,.
\end{equation}
In this case, the bounce action in the thick-wall approximation, in Eq.~(\ref{S3/Tn-handy}), gets shifted as
\begin{equation}
	\frac{S_3(T)}{T} \simeq \frac{430.72}{\sqrt{N_b}g_D^3}\frac{1}{\ln\frac{M}{T}}\sqrt{1\pm \left(\frac{T_0}{T}\right)^2} 
    \,, \label{S3/T-soft}
\end{equation}
where 
\begin{align} 
T_0 \equiv 2\sqrt{\frac{3}{N_b}}\frac{m_0}{g_D}\,. 
\end{align} 
Accordingly, $\beta_H$ in Eq.~(\ref{betaH}) takes the form 
\begin{equation}
	\label{deformedbeta}
	\beta_H \simeq - 4 + \frac{\sqrt{N_b}g_D^3}{430.72}\left[1\mp \left(\frac{T_0}{T_n}\right)^2\ln\frac{M}{T_n}\right]\left(\frac{S_3}{T_n}\right)^2.
\end{equation}
Thus a positive soft-scale breaking term generically tend to make $\beta_H$ smaller, while a negative one drives $\beta_H$ to increase. 
As we will see later, since the abundance of PBHs is highly sensitive to $\beta_H$, even a small soft-breaking term can lead to significant changes in the PBH production. 

We numerically evaluate $S_3(T)/T$ and $\beta_H$ as a function of $g_D$, $v_\phi$, and $m_0$. See Fig.~1, for some reference points in the parameter space. It is worth noting that the approximate formulae in Eqs.~(\ref{TnforCW}), (\ref{S3/T-soft}), and (\ref{deformedbeta}), are introduced merely to provide an intuitive understanding of the impact of the soft-scale breaking terms. The plot in this figure, as well as the subsequent results, are obtained from full numerical computations rather than relying on these approximations.


	

\begin{figure}[htbp!]
	\centering
	\includegraphics[scale=0.68]{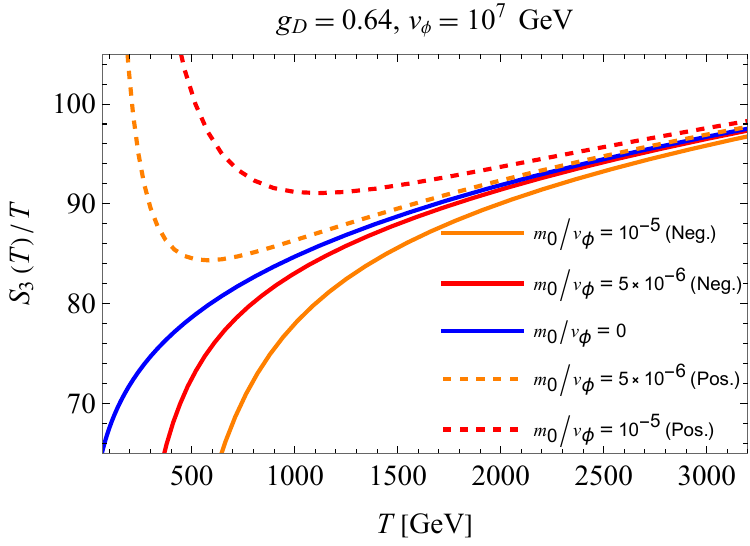}~~
	\includegraphics[scale=0.69]{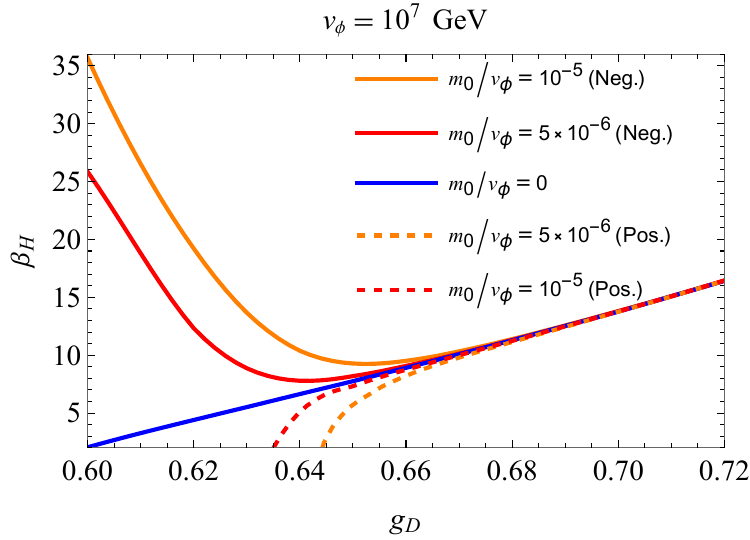}
	\caption{\textit{Left}: Plots of the bounce action $S_3(T)/T$ with a negative (denoted as ``Neg.") and positive (denoted as ``Pos.") soft-scale breaking terms, as a function of $T$, for reference values of $g_D$ and $v_\phi$. 
    \textit{Right}: Variations of $\beta_H$ with respect to the coupling $g_D$ for different mass choices and a fixed $v_\phi$ consistently with the PBH-dark matter interpretation (See Eq.~(\ref{vphi-range})). }  
	\label{beta}
	
\end{figure}

\section{Primordial Black Hole Formation}\label{sec:PBHs}

In this section we discuss PBH formation. As mentioned in the Introduction, due to the probabilistic nature of bubble nucleation, Hubble patches can be classified into two types: normal and delayed patches. Writing $t_c$ as the time corresponding to the critical temperature $T_c$, normal patches start to nucleate at $t_i = t_c$ where $t_i$ is the initial time, while delayed patches experience a delayed onset of nucleation at $t_i > t_c$.
The fraction of Hubble patches remaining in the false vacuum at time $t$ is then given by
\begin{equation}
F(t) = \left\{
\begin{array}{ll}
1, & \quad \text{if } t < t_i; \\
\displaystyle\exp\left\{-\frac{4\pi}{3}\int_{t_i}^{t} \mathrm{d}t' \Gamma(t') a^3(t') r^3(t,t')\right\}, & \quad \text{if } t > t_i.
\end{array}
\right.
\end{equation}
where $r(t, t')$ is the comoving radius of a bubble at $t$, which nucleates at $t'$:
\begin{align} 
r(t,t') = \int_{t'}^{t}\mathrm{d}t''\frac{v_w}{a(t'')}\,. 
\end{align} 
The percolation time is defined as the moment when the bubbles in the normal
patch form an infinite connecting cluster, i.e., $F(t_p) \simeq 0.71$~\cite{Rintoul:1997tze}. 
Furthermore, to ensure the completion of the supercooling phase transition, we impose the following criterion~\cite{Athron:2022mmm,Athron:2023mer,Turner:1992tz,Ellis:2018mja}: there must exist a temperature $T_e$ at which the fraction of the false vacuum drops below 1\%, and the physical volume of the false vacuum, $\mathcal{V}_\text{false} \propto a^3(t)F(t)$, is decreasing, i.e.,
\begin{equation}
\left. \frac{1}{\mathcal{V}_\text{false}} \frac{\mathrm{d} \mathcal{V}_\text{false}}{\mathrm{d}t} \right|_{T = T_e} < 0\,.
\end{equation}
If this condition is satisfied at $T_e$, the phase transition can be interpreted to be  successfully completed.

The evolution of the scale factor $a(t)$ is governed by the Friedmann equation 
with the total energy density given by 
the sum of the radiation ($\rho_r$) and the vacuum ($\rho_{v}$) energy densities,  
$\rho(t) = \rho_r(t) + \rho_v(t)$:  
\begin{equation}
	H(t) \equiv \frac{\dot{a}(t)}{a(t)} = \sqrt{\frac{1}{3M_{\rm pl}^2}\rho(t)}\,. 
\end{equation} 
The evolution of $\rho_r(t)$ follows the energy-momentum conservation in the expanding universe
\begin{equation}
	\frac{\mathrm{d}\rho_r(t)}{\mathrm{d}t} + 4H(t) \rho_r(t) = - \frac{\mathrm{d}\rho_v(t)}{\mathrm{d}t}\,.
\end{equation}
By taking into account the adiabatic time-temperature relation $\mathrm{d}t = -dT/\left(T H\right)$ during the phase transition, one can numerically solve the coupled equations for $\rho_r(t)$, $\rho_v(t)$, and $a(t)$. 
The vacuum energy density $\rho_v(t)$ in a given Hubble patch relates to the vacuum fraction via
\begin{equation}
    \rho_v(t) = F(t)\Delta V(t)\,,
\end{equation}
where $\Delta V(t)$ represents the energy density difference between the false and true vacua. During the supercooling stage, this difference remains nearly constant and can be well approximated by $V_0$ presented in Eq.~(\ref{Veffwithvacuum}).

In the expanding universe, the radiation energy density redshifts away as $\rho_r(t) \propto a^{-4}$, while the vacuum energy density $\rho_v(t)$ remains nearly constant. During the supercooling phase transition, as bubbles nucleate and grow, vacuum energy is converted into energy stored in the bubble wall, energy dumped into the plasma and sound waves from bubble collisions. Consequently, the universe evolves from being vacuum-dominated to radiation-dominated. Due to differences in transition times, delayed patches become overdense relative to surrounding normal patches. This overdensity is quantified as
\begin{equation}
	\label{threshold}
	\delta(t) \equiv \frac{|\rho_{\rm in}(t)-\rho_{\rm out}(t)|}{\rho_{\rm in}(t)}\,,
\end{equation}
where we have introduced the subscripts “in” and “out” to label delayed patches and normal ones, respectively, as well as their corresponding physical observables.
As the phase transition proceeds, $\delta(t)$ increases from 0 until the vacuum energy totally decays inside the overdense regions. At this moment, the overdensity reaches its maximum $\delta_{\rm max} = \delta(t_{\rm max})$. 
If $\delta_{\rm max}$ exceeds the threshold $\delta_c$, the delayed Hubble patch gravitationally collapses into a PBH. In the present study, we adopt 
$\delta_c=0.45$ for PBH formation~\cite{Musco:2004ak,Harada:2013epa}. 
This critical value is derived based on the assumption that the Universe is radiation-dominated and that overdense regions are spherically symmetric~\footnote{
In~\cite{Hashino:2025fse}, it has been shown that 
the parameter regions explored by PBH formation mechanisms 
are almost the same for the large density contrast and super-critical PBH production.  
}.


The probability $P_{\rm surv}(t_i)$ for null nucleation in the past light cone of a Hubble patch at $t_{\rm max}$ until $t_i$ is given by~\cite{Sato:1981gv,Kodama:1982sf}:
\begin{equation}\label{Psurv}
	P_{\rm surv}(t_i) 
	= \exp{\left[-\int_{t_c}^{t_i}\mathrm{d}t'\Gamma_{\rm in}(t')a_{\rm in}^3(t')V(t')\right]}\,,
\end{equation}
where the volume factor $V(t')$ is defined as~\cite{Kawana:2022olo,Gouttenoire:2023naa}
\begin{equation}
	V(t') = \frac{4\pi}{3}\left[\frac{1}{a_{\rm in}(t_{\rm max})H_{\rm in} (t_{\rm max})} + r(t_{\rm max},t')\right]^3\,.
\end{equation}
We define a critical delay time $t_i^{\rm PBH}$, representing the minimum delay of the onset of nucleation for a Hubble patch to reach the critical threshold at $t_{\rm max}^{\rm PBH}$, i.e., $\delta(t_{\rm max}^{\rm PBH})=\delta_c$.
In principle, any delayed patches with a nucleation time $t_i > t_i^{\rm PBH}$ can gravitationally collapse into a PBH, as a longer delay results in a larger overdensity. However, the probability $P_{\rm surv}(t_i)$ decreases rapidly as $t_i$ increases. Therefore, we conclude that the abundance of PBHs is primarily determined by the critical delay time. The probability that a Hubble patch collapses into a PBH is given by
\begin{equation}
    P_{\rm coll} \equiv P_{\rm surv}(t_i^{\rm PBH})\,.
\end{equation}
Moreover, it follows from Eq.~(\ref{Psurv}) that for the exponential nucleation, $P_{\rm coll}$ decreases exponentially with $\beta$, implying that a slower phase transition will lead to an enhancement in the PBH abundance.

The mass of the PBH is determined by the energy inside the sound horizon $c_sH^{-1}$ at the time of the collapse $t_{\rm max}^{\rm PBH}$~\cite{Escriva:2021pmf}
\begin{equation}
\label{PBH-mass}
	M_{\rm PBH} \simeq \frac{4\pi}{3} c_s^3 H_{\rm in}^{-3}(t_{\rm max}^{\rm PBH}) \rho_{\rm in}(t_{\rm max}^{\rm PBH}) = \frac{c_s^3}{2} \frac{M_{\rm pl}^2}{H_{\rm in}(t_{\rm max}^{\rm PBH})} \simeq M_{\text{\rightmoon}}\left(\frac{106.75}{g_\star(T_{\rm eq})}\right)^{1/2}\left(\frac{500\text{GeV}}{T_{\rm eq}}\right)^2,
\end{equation}
where $c_s = 1/\sqrt{3}$ is the sound speed during the radiation-dominated epoch, and $M_{\text{\rightmoon}} \simeq 3.7\times 10^{-8}M_\odot$ is the mass of the moon, with $M_\odot \simeq 1.99\times 10^{33}\text{ g}$ being the solar mass. $T_{\rm eq}$ is defined as the temperature at which the universe starts to be vacuum-dominated:
\begin{equation}
    \frac{\pi^2}{30} g_\star(T_{\rm eq}) T_{\text{eq}}^4 \equiv V_0 \,.
\end{equation}
If we regard PBH as the dark matter, the fraction of PBHs today is found to be
\begin{equation}
\label{PBH-abundance}
	f_{\rm PBH} \equiv \frac{\rho_{\rm PBH,0}}{\rho_{\rm DM,0}} = P_{\rm coll} \frac{M_{\rm PBH}\mathcal{N}_{\rm patches}}{\frac{4\pi}{3}H_0^{-3}\rho_{\rm DM,0}} \simeq \left(\frac{P_{\rm coll}}{6.2\times 10^{-12}}\right)\left(\frac{T_{\rm eq}}{500\text{GeV}}\right),
\end{equation}
where $\rho_{\rm DM,0}\simeq 0.26\times 3M_{\rm pl}^2H_0^2$ is the current dark matter energy density, and $\mathcal{N}_{\rm patches}$ represents the number of Hubble patches, when the Universe temperature was $T_{\rm eq}$, in our past light-cone, 
\begin{equation}
	\mathcal{N}_{\rm patches} = \left(\frac{a_{\rm eq}H_{\rm eq}}{a_0H_0}\right)^3 \simeq 5.3\times 10^{40} \times \left(\frac{g_\star(T_{\rm eq})}{100}\right)^{1/2}\left(\frac{T_{\rm eq}}{500\text{GeV}}\right)^3\,. 
\end{equation}
Here the ratio of the scale factor $a_{\rm eq}/a_0$ has been replaced as $a_{\rm eq}/{a_0} = \left(g_{\star S}(T_0)/g_{\star S}(T_{\rm eq})\right)^{1/3}(T_0/T_{\rm eq})$, which follows from the entropy conservation with the effective d.o.f. of entropy density at the present-day temperature $g_{\star S}(T_0)\simeq 3.94$.

According to~\cite{Gouttenoire:2023naa}, in a limit of ultra-supercooling $\alpha\geq 10^2$,  the fraction of causal patches collapsing into PBHs can be well-approximated by the analytic formula  
\begin{equation}
	\label{Pcoll}
	P_{\rm coll} \simeq \exp{\left[-a\left(\frac{\beta}{H_n}\right)^b (1+\delta_c)^{c\frac{\beta}{H_n}}\right]}\,,
\end{equation}
where the coefficients $a = 0.5646$, $b = 1.266$, and $c = 0.6639$ are obtained by  numerical fitting. From Eqs.~(\ref{PBH-mass}) and (\ref{PBH-abundance}), we infer that achieving $f_{\rm PBH}=1$ within the current CW-model framework requires a collapsed fraction in the range $10^{-16} \lesssim P_{\rm coll} \lesssim 10^{-13}$, which corresponds to $6.4 \lesssim \beta_H \lesssim 6.9$.
It is worth noting that the semi-analytical formula in Eq.~(\ref{Pcoll}) has been derived under the assumption of exponential nucleation during the supecooling phase transition, and becomes mathematically ill-defined for $\beta/H_n <0$. 
This feature likely indicates a breakdown of the exponential approximation in the presence of a positive soft-scale breaking mass term ($m_0 \neq 0$), in which case a nucleation approximation based on the second-order expansion of the bounce action may provide a more accurate description. This issue will be discussed in detail in the following section.

For PBHs that can explain the whole component of dark matter today, 
the mass needs to lie in the range of~\cite{Carr:2020gox} 
\begin{equation}
    \label{asteroid-mass}
    10^{-16}M_\odot\lesssim M_{\rm PBH} \lesssim 10^{-10}M_\odot\,.
\end{equation}
This constraint can be rephrased in terms of an equilibrium temperature $T_{\rm eq}$ in the range of 
$5\times10^{4}\text{ GeV} \lesssim T_{\rm eq} \lesssim 10^{7}\text{ GeV}$,
or equivalently, the $U(1)_D$ symmetry breaking scale in the range of 
\begin{align}
3\times10^5\text{ GeV} \lesssim v_\phi \lesssim 3\times10^8\text{ GeV}
\,.\label{vphi-range}
\end{align} 

In the absence of the soft-scale breaking mass, $m_0=0$, each value of $v_\phi$ within the range in Eq.~(\ref{vphi-range}) corresponds to a unique gauge coupling $g_D$ that yields the required PBH abundance with $f_{\rm PBH} = 1$. However, when $m_0 \neq 0$, for a given $v_\phi$, the the requirement can then be satisfied over a finite interval of $g_D$, depending on the sign and magnitude of $m_0$.
This broadens the viable parameter space in CW-type models and enhances their ability to account for dark matter through PBH production. 


\section{Impact of soft-scale breaking mass on PBH Abundance}\label{sec:Significance}
In this section, we discuss the validity of the exponential approximation of the nucleation rate (\ref{nucleation-rate}) in the CW-type supercooling. 
To this end, we perform a second-order Taylor expansion of $S(T) \equiv S_3(T)/T - 4 \ln T$ around the nucleation time $t_n$, to get 
\begin{equation}
	S(t) \simeq S(t_n) -\beta (t-t_n) + \frac{\beta_V^2}{2}(t-t_n)^2 \,.
\end{equation}
The nucleation rate then takes the form
\begin{equation}
	\Gamma(t) \simeq 
    \Gamma_n\, \mathrm{e}^{\beta (t-t_n) - \frac{\beta_V^2}{2}(t-t_n)^2}\,,  
\end{equation}
with
\begin{align}
	\frac{\beta}{H_n} &\simeq -4 + T_n \left.\frac{\mathrm{d}(S_3/T)}{\mathrm{d}T}\right|_{T=T_n}\,, \notag \\ 
	\frac{\beta_V}{H_n}
    &\simeq \left.\sqrt{T \frac{\mathrm{d}(S_3/T)}{\mathrm{d}T}  + T^2\frac{\mathrm{d^2}(S_3/T)}{\mathrm{d}T^2}}\right|_{T=T_n}\,, \notag\\ 
    \Gamma_n &\simeq T_n^4\, \mathrm{e}^{-S_3(T_n)/T_n}\,. 
\end{align}
Here $H_n \equiv H(T_n)$ and terms involving $H'(T)\equiv \mathrm{d}H(T)/\mathrm{d}T$ have been neglected, which works well for the ultra-supercooling epoch. 
\begin{figure}[htbp!]
	\centering
	\includegraphics[scale=0.68]{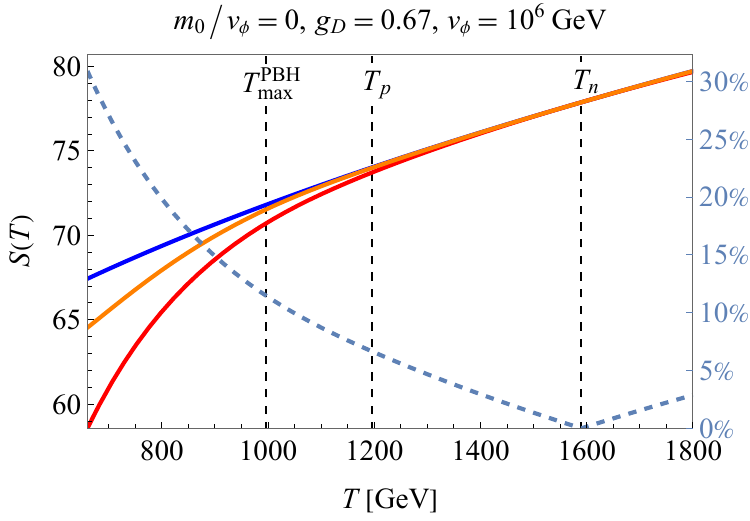}
	\includegraphics[scale=0.698]{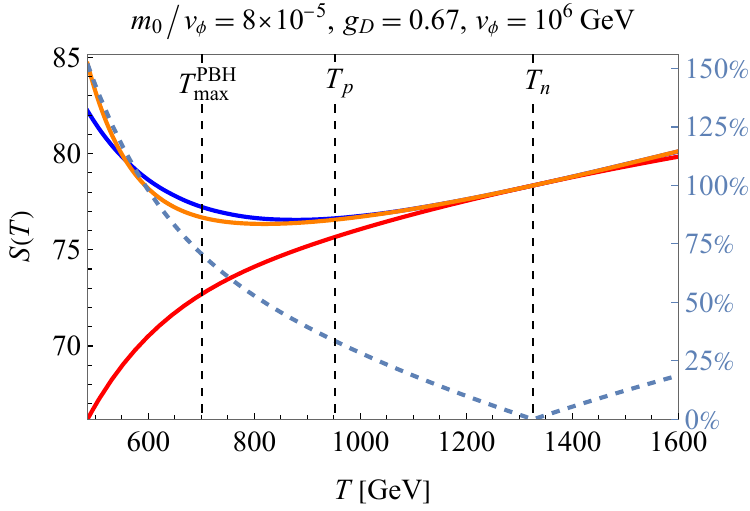}
	\caption{Plots of $S(T)\equiv S_3(T)/T - 4\ln T $ as a function of $T$ for two benchmark parameter sets. The blue, red, and orange curves respectively represent the result without any approximation on the nucleation, with linear approximation, and the second-order approximation. The vertical dashed lines (from right to left) indicate the nucleation temperature $T_n$, percolation temperature $T_p$, and PBH formation time $T_{\rm max}^{\rm PBH}$. Also have been displayed the relative error $\epsilon_\beta$ in Eq.~(\ref{linearity}) (in unit of percent), which have been drwan by the dashed-light blue curves.}
	\label{action-approx}
\end{figure}
The linear approximation of $S(t)$, or equivalently, exponential nucleation approximation of $\Gamma(t)$ is valid if the quadratic term is much smaller than the linear term, i.e., the relative error is small enough as follows: 
\begin{equation}
    \epsilon_\beta(t) \equiv \frac{\beta_V^2(t-t_n)}{2\beta} \ll 1\,.
\label{linearity}
\end{equation}


Given the generic expressions as above, we now discuss the effect of nonzero $m_0$ 
on the dynamics of the ultra-supercooling phase transition. 

\begin{enumerate}
\item \textbf{Case 1 with $m_0 = 0$}:

A typical feature in this case is that the bounce-action curve exhibits a tangent-like behavior, indicating the ultra-supercooling phase transition at sufficiently low temperatures, as long as $g_D$ is not too small. 
In Table~\ref{tab:table1}, based on numerical calculations (not using approximated formulae), we show that  
for a fixed $v_\phi$, $\beta_H$ decreases as $g_D$ decreases. 
This feature, in turn, leads to a substantial enhancement in the abundance of PBHs.
The linear expansion, i.e. applying the condition in Eq.~(\ref{linearity}), is reliable for evaluating the bounce action up to the point of PBH formation;  
for instance, the relative error stays below 10\% for a benchmark point $g_D = 0.67, v_\phi = 10^6 \text{ GeV}$. 
Thereby, the PBH abundance can be estimated using the exponential nucleation approximation. 

\item \textbf{Case 2 with negative soft-scale breaking $-m_0^2\phi^2$}: 

This scenario is similar to Case 1, but with a negative squared-mass term that lowers the barrier between the two vacua and results in a steeper tangent-like action curve. This leads to an increase in $\beta_H$, causing a faster phase transition compared to Case 1. Consequently, the resulting PBH abundance is suppressed due to the sharp drop in $P_{\rm coll}$. This trend may revive models of CW-type which used to produce overabundance of PBHs with $m_0=0$.
As seen from the reference points listed in Table~\ref{tab:table2}, 
even a tiny variation in $m_0/v_\phi$ at the level of $10^{-6}$ can shift the PBH abundance by many orders of magnitude from an overproduction regime into the observationally allowed region. 

\item \textbf{Case 3 with positive soft-scale breaking $+m_0^2\phi^2$}:

In this case, the potential barrier persists even at $T \sim 0$, so that 
the bounce-action curve exhibits a skewed U-shaped behavior. 
If $m_0$ is sufficiently small, $T_n$ and $T_p$ remain far from $T_{\rm min}$ at which $S_3/T$ reaches the minimum. 
This feature validates the exponential nucleation approximation during the phase transition, allowing to reliably predict the PBH abundance. 
However, when $m_0$ gets larger and larger, 
$T_n$ approaches $T_{\rm min}$, hence the exponential approximation is to be broken  down. As shown in the right panel of Fig.~\ref{action-approx}, the relative error size in Eq.~(\ref{linearity}) can reach 70\% at the time of PBH formation. 
The thus non-negligible Gaussian damping of the nucleation rate leads to a slower phase transition, hence the PBH abundance gets amplified. 
Table~\ref{tab:table3} illustrates that for even a tiny $m_0$, like $ m_0/v_\phi = 2.5\times 10^{-5}$, the PBH abundance $f_{\rm PBH}$ can be enhanced by 16 orders of magnitude (for $g_D=0.67, v_\phi=10^7\text{ GeV}$), compared to the case with $m_0=0$.
This makes it possible to fully explain the dark matter abundance today by PBHs. 
The right panel of Fig.~\ref{action-approx} also shows that 
the second order contribution (orange curve) beyond the linear approximation (red curve) is essential to provide the actual bubble nucleation (blue curve).  
\end{enumerate}
\begin{table}[htbp!]
\caption{\label{tab:table1}%
The estimated numbers for the PBH mass ($M_{\rm PBH}$) and the PBH-dark matter fraction ($f_{\rm PBH}$ defined as in Eq.(\ref{PBH-abundance}))  in the case with $m_0=0$ 
for varying values of $g_D$ at a fixed $v_\phi = 10^7\text{ GeV}$, listed 
together with $T_n$, $T_p$, $\beta_H$.  
}
\begin{ruledtabular}
\begin{tabular}{cccccc}
$g_D$ & $T_n\,[{\rm GeV}]$ &  $T_p\,[{\rm GeV}]$ & $\beta_H$ & $M_{\rm PBH}$ & $f_{\rm PBH}$\\
\colrule
0.63 & 875.76 & 630.78 & 5.52 & $2.83\times10^{-14}M_\odot$ & $1.58\times 10^{9}$ \\
0.64 & 1690.44 & 1212.77 & 6.64  & $2.75\times10^{-14}M_\odot$ & $1.15\times 10^{6}$ \\
0.65 & 2978.28 & 2141.82 & 7.76  & $2.66\times10^{-14}M_\odot$ & $2.34\times 10^{1}$ \\
0.66 & 4889.61 & 3536.68 & 8.90 & $2.58\times10^{-14}M_\odot$ & $2.00\times 10^{-6}$ \\
0.67 & 7587.35 & 5529.61 & 10.06  & $2.51\times10^{-14}M_\odot$ & $1.01\times 10^{-16}$ \\
0.68 & 11242.30 & 8263.42 & 11.29 & $2.43\times10^{-14}M_\odot$ & $8.09\times 10^{-32}$ \\
\end{tabular}
\end{ruledtabular}
\end{table}
\begin{table}[htbp!]
\caption{\label{tab:table2}%
The estimated numbers for $f_{\rm PBH}$ in the case with a negative soft-scale breaking mass term $m_0$: $-m_0^2\phi^2$,  
for $g_D = 0.64$ and $v_\phi = 10^7\text{ GeV}$, listed 
together with $T_n$, $T_p$, $\beta_H$.  
Here we have $M_{\rm PBH}\simeq2.75\times 10^{-14}M_\odot$ in all cases. 
}%
\begin{ruledtabular}
\begin{tabular}{ccccc}
$m_0/v_\phi$    &  $T_n\,[{\rm GeV}]$ &  $T_p\,[{\rm GeV}]$ & $\beta_H$&  $f_{\rm PBH}$\\
\colrule
0 & 1690.44 & 1212.77 & 6.64 &  $ 1.15\times 10^{6}$ \\
$2\times 10^{-6}$ & 1709.28 & 1227.1 & $6.85$ &  $1.94\times 10^{5}$ \\
$4\times 10^{-6}$ & 1762.26  & 1268.1  & $7.41$ & $3.61\times 10^{2}$ \\
$6\times 10^{-6}$ & 1842.55  & 1331.76  & 8.28 &  $6.11\times10^{-4}$ \\
$8\times 10^{-6}$ & 1942.39  & 1412.81  & 9.27 &  $4.38\times 10^{-15}$ \\
$10\times 10^{-6}$ &  2055.35 &  1506.42 & 10.41 &  $2.37\times 10^{-35}$ \\
\end{tabular}
\end{ruledtabular}
\end{table}
\begin{table}[htbp!]
\caption{\label{tab:table3}%
The estimated numbers for $f_{\rm PBH}$ in the case with a positive soft-scale breaking mass term $m_0$: $+m_0^2\phi^2$,  
for $g_D = 0.67$ and $v_\phi = 10^7\text{ GeV}$, listed 
together with $T_n$, $T_p$, $\beta_H$.   
Here we have $M_{\rm PBH}\simeq2.51\times 10^{-14}M_\odot$ in all cases. 
}
\begin{ruledtabular}
\begin{tabular}{ccccc}
	$m_0/v_\phi$ &
	$T_n\,[{\rm GeV}]$ &
	$T_p\,[{\rm GeV}]$ &
	$\beta_H$&
	$f_{\rm PBH}$\\
	\colrule
	0 & 7587.35 & 5529.61 & 10.06 &  $ 1.01\times 10^{-16}$ \\
	$0.5\times 10^{-5}$ & 7567.94 & 5512.56 & $9.99$ &  $9.63\times 10^{-16}$ \\
	$1.0\times 10^{-5}$ & 7508.90  & 5460.82  & $9.79$ & $2.58\times 10^{-13}$ \\
	$1.5\times 10^{-5}$ & 7407.63  & 5372.43  & 9.43 &  $1.75\times10^{-9}$ \\
	$2.0\times 10^{-5}$ & 7259.19  & 5243.79  & 8.90 &  $7.99\times 10^{-5}$ \\
	$2.5\times 10^{-5}$ &  7055.00 &  5068.97 & 8.14 &  $5.71\times 10^{0}$ \\
\end{tabular}
\end{ruledtabular}
\end{table}

\section{Reference viable models of CW-type with $m_0^2 \neq 0$} 

In this section, we address two scenarios of CW-type ultra-supercooling having a soft-scale breaking mass $m_0^2\neq 0$ with definitely positive (Sec.~\ref{sec:walking-model}) and negative (Sec.~\ref{sec:B-L-model}) signs, and those related explicit predictions to the PBH dark matter production.  

\subsection{A scenario to achieve positive soft-scale breaking}\label{sec:walking-model}

A simple-minded dilaton-portal modeling 
would be trivial to provide a negative mass-squared for the dilaton 
 by allowing 
the coupling to another scalar field, say, $S$: 
$V= - \kappa S^2 |\phi|^2 + \lambda_S S^4$ with $\kappa>0$ and $\lambda_S>0$, 
which gives $\langle S \rangle \neq 0$ along the flat direction. 
However, it would be somewhat nontrivial 
to supply a positive-mass squared term 
into the CW type potential: 
to achieve $\langle S \rangle \neq 0$ through 
the portal interaction, $\kappa<0$ is required, not $\kappa>0$.


One candidate to realize a positive portal coupling 
to $\phi$ with nonzero vacuum expectation value (VEV) of $S$  is the case 
when $\phi$ is a composite dilaton arising from 
a dark QCD theory, in a manner like what 
has been discussed in the literature~\cite{Ishida:2019wkd,Cacciapaglia:2023kat,Zhang:2023acu,Zhang:2024vpp,Liu:2024xrh}. 
The key point is to notice the emergence of 
the so-called chiral chemical potential around 
the dark QCD phase transition epoch at $T \sim \Lambda_d$. 
In hot dark-QCD plasma a local CP-odd domain may be created due to the presence of the sphaleron as in the case of QCD~\cite{Manton:1983nd,Klinkhamer:1984di}, so that the QCD vacuum characterized by 
the strong CP phase $\theta$ and its fluctuation (in the spatial-homogeneous direction) gets significantly sizable~\cite{Kharzeev:2007tn,Kharzeev:2007jp,Fukushima:2008xe} within the QCD time scale~\cite{McLerran:1990de,Moore:1997im,Moore:1999fs,Bodeker:1999gx}: 
the sphaleron transition rate is not suppressed by the thermal effect in contrast to 
the instanton's one~\cite{Moore:1997im,Moore:1999fs,Bodeker:1999gx}. 
The time fluctuation $\partial_t \theta(t)$, to be referred to as the chiral chemical potential, $\mu_5$~\cite{Kharzeev:2007tn,Kharzeev:2007jp,Fukushima:2008xe,Andrianov:2012hq,Andrianov:2012dj}, will be significant as well when the non-conservation law of the $U(1)$ axial symmetry is addressed.

For this hot-dark QCD plasma, a pseudoscalar $S$ is also assumed to couple through coupling to the $N_f$ flavor Dirac dark quarks $(F_{L,R}^{i})$ ($i=1,\cdots, N_f$) in a $U(1)$ axial-, the dark QCD-gauge-, and scale-invariant way via 
the Yukawa coupling term, 
\begin{align}
   i y_S  \bar{F}_L S F_R + {\rm h.c.}
\,.  \label{yS}
\end{align}
Then, the $S$ coupling to the chiral chemical potential 
would also be induced through the axial covariance: 
\begin{align}
    |D_\mu S|^2 \,, 
    \qquad {\rm with} \qquad 
    D_\mu = \partial_\mu - i \mu_5 \delta^0_\mu 
    \,. 
\end{align}
Hence $S$ gets the effective potential along with $\mu_5$ 
\begin{align} 
  V_{\mu_5}(S) = - \frac{\mu_5^2}{2} |S|^2
\,. 
\end{align}
This is definitely a negative mass-squared term ensured by the gauge covariance for the axialvector field $A_\mu = \mu_5 \delta_\mu^0$, which corresponds to 
the repulsive force from the vector interaction against the scalar probes charged under the corresponding gauge. 
This is a dynamical scale-explicit breaking generated by 
the dark QCD thermal plasma. 
Thus, combined with the quartic $S^4$ term, $S$ can get 
nonzero VEV, $\langle S \rangle = \sqrt{\frac{\mu_5^2}{\lambda_S}}$. 
Then the portal coupling to $\phi$, 
$V_{S^2\phi^2} = + \kappa  |S|^2 |\phi|^2$ 
with $\kappa>0$ precisely provides a positive mass-squared term for $\phi$: 
\begin{align}
V_{m_0^2} = m_0^2 |\phi|^2 
  \,, \qquad {\rm with} \qquad 
  m_0^2 = \frac{\kappa \mu_5^2}{\lambda_S}  
    \,.  \label{mu5-mass}
\end{align}
As long as the dark QCD sector and $S$ are 
in the common thermal bath,
the $\mu_5$ contribution above 
can survive with the related sphaleron transition rate $\sim \alpha_d^5 T^4$~\cite{McLerran:1990de}. 
Even if the theory undergoes supercooling, 
the local thermal plasma is still hot enough 
to keep the sphaleron rate transition comparable with 
the accumulated potential energy $V = V_{\rm thermal} \sim T^4$~\footnote{
One might think about generating a suitable positive breaking term from a conformal coupling of the dilaton to the Ricci scalar. However, the large hierarchy between the Planck scale and the dilaton scale significantly suppresses the induced mass term, yielding $m_0/v_\phi \sim \sqrt{\xi_\phi}\,v_\phi/M_{\rm pl}$, 
where $\xi_\phi$ is the non-minimal coupling constant. This ratio is typically several orders of magnitude below the benchmark value of $\mathcal{O}(10^{-6})$ suggested by our analysis. As a result, this gravitational mechanism alone cannot account for the necessary deformation, and additional ultraviolet sources of explicit scale symmetry breaking must be invoked.
}.

The size of $m_0^2$ in Eq.~(\ref{mu5-mass}) can be parametrically 
smaller than $v_\phi^2$, when dark QCD is like a walking gauge theory with many flavors (e.g., $N_f=8$~\cite{Aoki:2013xza,LSD:2014nmn,Hasenfratz:2014rna}), 
because it is anticipated that $\mu_5 = {\cal O}(T_c)$ for the chiral-scale phase transition in the walking 
QCD, which is $\ll v_\phi$ as addressed in the literature~\cite{Ishida:2019wkd,Cacciapaglia:2023kat,Zhang:2023acu,Zhang:2024vpp,Liu:2024xrh}. 
The portal coupling $\kappa$ is also supposed to be small enough, because 
the dark QCD sector is external to $S$: the expected amplitude would go like 
$\sim {\cal O}(g_{\phi FF}^2 y_s^2 N_{C}/(4\pi)^2)$ with $g_{\phi FF} \sim m_F/f_\phi$ 
and $N_C$ being the number of dark QCD colors. 
Thus at any rate, one can realize $m_0 \ll v_\phi$.

Thermalization of this walking dark QCD sector with the SM plasma 
can be ensured by gauging the dark quark-chiral symmetry in part by the electroweak 
and/or $U(1)_{B-L}$ gauges, as have been discussed in the literature~\cite{Ishida:2019wkd,Liu:2024xrh}. 
Thus the walking dilaton $\phi$ can, at the beginning, be thermally stabilized at $\phi=0$ in the potential (or at around there due to the walking pion mass~\cite{Zhang:2023acu}) and will be trapped there until the potential barrier is gone.

To make this scenario more successful, the size of the Yukawa coupling $y_S$ also 
needs to be small enough. 
Otherwise $S$ can develop a sizable tadpole potential term, i.e., trigger 
generation of the VEV, via the mixing with the dark $\eta'$ meson ($\eta_d'\sim \bar{F}i \gamma_5 F$): $V_{\rm tad} \sim - 2 y_S \Lambda_d^2 {\rm Re}[S] \eta'_d$.

The phase part of the $S$ field may act as an axionlike particle, and can decay into diphoton via the coupling to the SM, leaving astrophysical and collider experimental signals, which depend on the mass scale. Thus this scenario could also provide a detectable phenomenological consequence. 
Detailed analyses of such experimental signals are beyond the scope of the present work, 
to be explored elsewhere.

In the following, we adapt the linear sigma model with 
$U(N_f)_L\times U(N_f)_R$ chiral symmetry as an effective theory of the underlying dark QCD scenario. The one-loop effective potential at finite temperature including daisy diagrams in the Parwani scheme~\cite{Parwani:1991gq}, is given by
\begin{equation}
V_{\rm eff}(\phi,T) 
=  \frac{m_0^2}{2}\phi^2 + \frac{N_f^2-1}{64\pi^2} 
\mathcal{M}_{s^i}^4(\phi, T)
\left( \ln{\frac{
		\mathcal{M}_{s^i}^2(\phi, T)
	}{\mu_{\rm GW}^2}} - \frac{3}{2} \right) 
+\frac{T^4}{2\pi^2} (N_f^2-1) J_B\left(
\mathcal{M}_{s^i}^2(\phi, T)
/T^2\right)\,,
\label{eq:veffT}
\end{equation}
where $(s^0\equiv \phi,\, s^i)$ represent the $N_f^2$ scalars with quantum numbers $J^{PC}=0^{++}$, while the remaining $N_f^2$ pseudoscalars $(p^0,\,p^i)$, associated with  $J^{PC}=0^{-+}$  are absent from the potential due to their vanishing $s^0$-dependent mass, as required by classical scale invariance. The masses of adjoint representation scalars  $m_{s^i}^2(\phi)\equiv \frac{2\lambda_2}{N_f}\phi^2$ have been dressed as 
$\mathcal{M}_{s^i}^2(\phi, T) = m_{s^i}^2(\phi) + \Pi(T)$ with the Debye mass
\begin{align}
\Pi(T) = \frac{T^2}{6}\bigl((N_f^2 + 1)\lambda_1 + 2N_f \lambda_2\bigr)\Big|_{\lambda_1 = -\lambda_2 / N_f} \,.
\end{align}
Detailed supplements to the linear sigma model and the computation of thermal corrections within the present model are provided in Appendix~\ref{LSM}. It is worth noting that this model differs slightly from the conventional classically scale-invariant case: due to the large number of degrees of freedom, $\mathcal{O}(N_f^2)$ with $N_f = 8$ in the case of many flavor walking, the thermal mass becomes approximately at more than ten times larger than the field-dependent mass near the barrier. As a result, it is necessary to include the daisy (ring) resummation contributions.

\begin{figure}[htbp!]
	\centering
	\includegraphics[scale=0.68]{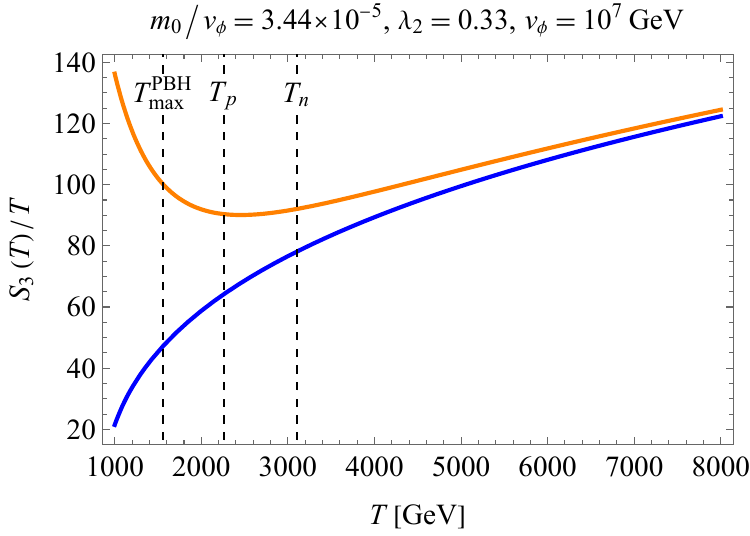}
	\includegraphics[scale=0.63]{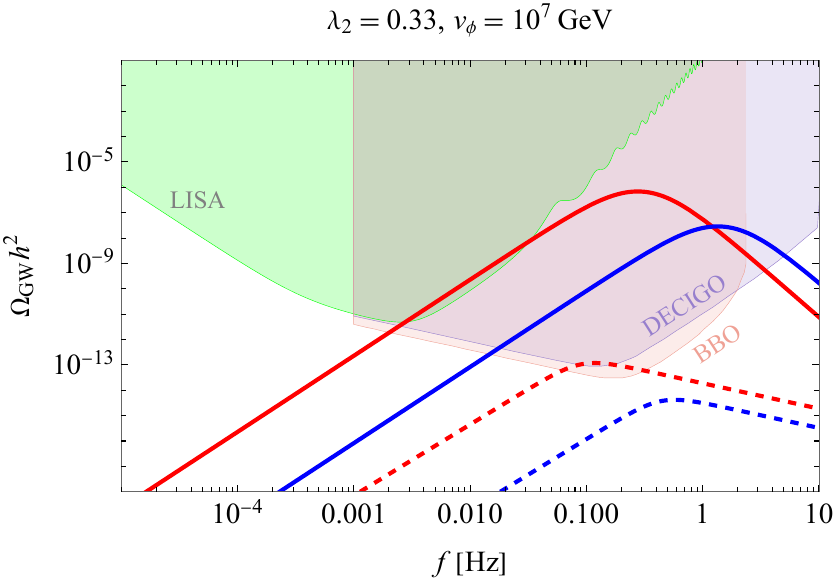}
	\caption{\textit{Left}: Plots of $S_3(T)/T$ as a function of $T$ for two benchmark parameter sets, BP A (blue) and BP A$'$ (Orange). The vertical dashed lines (from right to left) indicate the nucleation temperature $T_n$, percolation temperature $T_p$, and PBH formation time $T_{\rm max}^{\rm PBH}$ for BP A$'$. \textit{Right}: Stochastic GW spectra in the cases with (red) and without (blue) the soft-scale breaking term. For each case, the solid and dashed curves correspond to the spectra sourced from sound waves and bubble wall collisions, respectively. 
    The GW signals are compared with future prospected detector sensitivities~\cite{Moore:2014lga,Schmitz:2020syl}. } 
	\label{walking-figure}
\end{figure}

To illustrate the PBH enhancement mechanism within this benchmark model framework, we numerically compute the phase transition dynamics using the benchmark parameters $(\lambda_2,v_\phi) = (0.33,10^7\text{ GeV})$, with and without the positive soft-scale breaking mass term.  
As shown in the first row denoted as ``BP A" of Tab.~\ref{tab:walking}, we can see that the case with $m_0=0$ yields an ultra-supercooling, and the phase transition proceeds so rapidly that no PBH is produced. The corresponding bounce action $S(T)$ exhibits a tangent-like profile, as illustrated by the blue curve in the left panel of Fig.~\ref{walking-figure}. However, when the positive-soft scale breaking term as in Eq.(\ref{mu5-mass}) is included, the bounce action develops a skewed U-shaped behavior, as expected from the generic discussion in Sec.~\ref{sec:Significance} (See the second row in labeled as ``BP A$^\prime$" of Tab.~\ref{tab:walking}). This characteristics is reflected in the duration of the phase transition, hence enhances the production of PBHs to reach $f_{\rm PBH} \sim 1$. 
In Fig.~\ref{walking-figure}, also has been plotted the GW spectra predicted from 
the cases with (red) and without (blue) the soft-scale breaking mass term. 
The detailed discussions on GW spectra are presented in Appendix~\ref{sec:GWs}.  
\begin{table}[htbp!]
	\caption{\label{tab:walking}%
		Predicted parameters related to the supercooling phase transition and PBH production for two benchmark points (BPs) for the many-flavor QCD scenario with $N_f=8$. 
	}
	\begin{ruledtabular}
		\begin{tabular}{lc|cccccc
        }
			 & $m_0/v_\phi$ & $T_n[{\rm GeV}]$ & $T_p[{\rm GeV}]$ &
			$\alpha$ &
			$\beta_H$& 
			$M_{\rm PBH}$ &
			$f_{\rm PBH}$\\
			\colrule
			\text{BP A} & 0 & $4272.8$ & $3598$ &  $6.38\times 10^{7}$ & $40.55$ 
            & $2.975\times 10^{-14}M_\odot$ & $ \sim 0$ \\
			BP A$'$ & $3.44\times 10^{-5}$ & $3105.1$ & $2257.4$ & $4.12\times 10^{8}$ & $11.75$ 
            & $2.975\times 10^{-14}M_\odot$ & $\sim1$
		\end{tabular}
	\end{ruledtabular}
\end{table}
\if{
Under the above parameter settings, the ratio of Lorentz boosts of the bubble wall at collision to that at equilibrium, $\gamma_\star/\gamma_{\rm eq}>1$, indicates that the bubbles reach their terminal velocity before colliding, regardless of whether the breaking term is present. Consequently, most of the vacuum energy releases into the surrounding plasma rather than the bubble walls, as quantified by the energy partition ratio $\kappa_{\rm sw}/\kappa_{\rm coll}\sim \mathcal{O}(10^2)$, and sound waves become the dominant source in the GW spectrum. As demonstrated in the right-hand panel of Fig.~\ref{walking-figure}, the peak frequency amplitudes of GW spectrum sourced by sound waves fall within the sensitivity bands of space-based GW detectors such as LISA, DECIGO, and BBO, whereas the peak frequencies from bubble collisions lie outside their detection ranges. The presence of a mass-squared breaking term introduces an additional potential barrier, which lowers the peak frequency by slowing down the phase transition, and enhances the spectral amplitude due to a larger $\kappa_{\rm sw}$.
}\fi

\subsection{An illustrative example for negative-soft scale breaking}\label{sec:B-L-model}
We consider the classically conformal $U(1)_{B-L}$ extension of the SM as a reference model having a negative soft-scale breaking term. 
This is a concrete example of the Higgs-dilaton portal scenario which has briefly been 
referred to in the previous subsection. 
In this framework, the $B-L$ symmetry in the SM is promoted to a local gauge symmetry along with a the gauge field $Z^\prime$. 
The SM quarks and leptons carry $U(1)_{B-L}$ charges $+1/3$ and $-1$, respectively, while the SM Higgs doublet transforms as a singlet under $U(1)_{B-L}$.
To achieve spontaneous breaking of the $B-L$ gauge symmetry and to ensure cancellation of gauge anomalies~\cite{Iso:2009ss,Iso:2009nw,Jinno:2016knw}, we introduce a $B-L$ gauged scalar $\Phi$ with the $B-L$ charge $Q_{B-L}=-2$, as well as three right-handed neutrinos $N_R^{(i=1,2,3)}$ 
with $Q_{B-L}=-1$ for each. 
The classically scale-invariant tree-level scalar potential involving the Higgs doublet $H$ and the $B-L$ Higgs $\Phi$ is given as  
\begin{equation}
    V(H,\Phi) = \lambda_h |H|^4 + \lambda_\phi |\Phi|^4 - \lambda_{h\phi}|H|^2|\Phi|^2\,. 
\end{equation}
The minus sign in the portal coupling term ($\lambda_{h\phi} |H|^2 |\Phi|^2$) ensures that electroweak symmetry is spontaneously broken after the $B-L$ gauge symmetry breaking, giving rise to the Higgs mass $m_h = \sqrt{\lambda_{h\phi}}v_\phi=125\text{ GeV}$. 
Following the approach of Gildener and Weinberg~\cite{Gildener:1976ih}, the tree-level potential exhibits a flat direction at a particular scale $\mu_{\rm GW}$. At the one-loop level, the flat direction will be lifted by quantum corrections, leading a CW type potential. This model can directly be projected onto the one that we have introduced as a dark $U(1)_{D}$ model in Secs~\ref{sec:model} and~\ref{sec:PT}, when $g_{B-L} \gg \lambda_{h},  \lambda_{\phi}$ with $g_{B-L} \equiv g_{D}/2$. 
This relatively large coupling $g_{\rm B-L}$ also ensures 
the thermalization of the $B-L$ sector with the SM plasma in the early Universe. 
\if{
\begin{equation}
    V_{\rm CW}^{B-L}(\phi) = \frac{\beta_{\lambda\phi}}{4}\phi^4\left[\ln\frac{\phi}{v_\phi}-\frac{1}{4}\right]\,,
\end{equation}
where
\begin{equation}
    \beta_{\lambda\phi} \equiv \frac{\mathrm{d}\lambda_\phi}{\mathrm{d}\ln\mu} = \frac{1}{(4\pi)^2}\left[96g_{B-L}^4+20\lambda_\phi^2-(48g_{B-L}^2-6Y^2)\lambda_\phi-\sum_i Y_i^4\right]\,,
\end{equation}
and $v_\phi = \sqrt{2}\langle\Phi\rangle$ is the VEV determined at the scale $\mu_{\rm GW}$ by the stationary condition.
We compute the one-loop effective potential and see that the dominant contribution arises from the gauge boson $Z^\prime$, allowing us to neglect other field-dependent corrections in the effective theory. This leads to a simplified expression for the thermal effective potential
}\fi 
Thus the one-loop effective potential at finite $T$, including only the dominant 
$U(1)_{B-L}$ gauge loop contribution, is evaluated in a way as done in Sec.~\ref{sec:model}, to be  
\begin{equation}
    V^{B-L}_{\rm eff}(\phi,T) \simeq \frac{\beta_{\lambda\phi}}{4}\phi^4\left[\ln{\left(\frac{\phi}{v_\phi}\right)}-\frac{1}{4}\right] + \frac{3T^4}{2\pi^2}J_B\left(\frac{m_{Z'}^2}{T^2}\right) + \frac{T}{12\pi}\left[m_{Z'}^3-\left(m_{Z'}^2+\Pi_{Z'}\right)^{3/2}\right]\,,
\end{equation}
where  $v_\phi = \sqrt{2}\langle\Phi\rangle$; 
$\beta_{\lambda_\phi} = \frac{1}{(4\pi)^2}\left(96g_{B-L}^4 
\right)$; 
$m_{Z'} = 2 g_{B-L}\phi$ is the field-dependent mass of the $Z^\prime$ boson; $\Pi_{Z'} = 4g_{B-L}^2T^2$ its Debye mass.

The present $B-L$ symmetry-phase transition predicts a percolation temperature (much) below the QCD critical temperature, $T_{\rm QCD}\sim 100\text{ MeV}$, much like QCD-driven electroweak-phase transition scenarios that have been extensively discussed in the literature~\cite{Iso:2017uuu,vonHarling:2017yew,Marzo:2018nov,Bodeker:2021mcj,Sagunski:2023ynd,Gouttenoire:2023pxh,Guan:2024ccw}. 
Consequently, in prior to the electroweak phase transition, 
 the QCD-driven quark condensates induces a small VEV on the QCD scale for the Higgs field,  which is dominated by the top Yukawa interaction~\cite{Witten:1980ez}, 
\begin{equation} 
    \langle h \rangle = v_{\rm QCD} \sim \left(y_t\langle\bar{t}t\rangle/(\sqrt{2}\lambda_h)\right)^{1/3} \sim \Lambda_{\rm QCD}\,. 
\end{equation}
This generates a negative soft-scale breaking mass for the $B-L$ Higgs $\phi$ 
at $T \lesssim T_{\rm QCD}$ via the Higgs portal $\lambda_{h \phi}$ coupling term: 
\begin{equation}
    -\frac{1}{4}\lambda_{h\phi}v_{\rm QCD}^2\phi^2\,.
\end{equation}
After supercooling phase transitions for the $B-L$ and electroweak symmetries are completed at $T=T_p \ll T_{\rm QCD}$, the Higgs field acquires the desired VEV $\simeq 246$ GeV.

The numerical analysis on the phase transition dynamics, PBH, and GW productions 
is summarized in Table~\ref{tab:B-L}, in comparison with the case having no soft-scale breaking mass (labeled as ``BP B"). 
There we have taken $(g_{B-L},v_\phi$) = ($0.273, 4.3\text{ TeV}$),  
which leads to $\lambda_{h\phi} \simeq 8.5 \times 10^{-4}$ and $m_0/v_\phi = \sqrt{\lambda_{h\phi}/2} \cdot (v_{\rm QCD}/v_\phi) \simeq 5.4 \times 10^{-7}$ for $v_{\rm QCD} = 100$ MeV 
(labeled as ``BP B$^\prime$"). 
The GW spectra are dominated by sound waves 
and the peak frequency and amplitude are estimated to be ($f_{\rm sw}^{\rm peak}=0.1\text{ mHz}$, $\Omega_{\rm sw}^{\rm peak}h^2=5.9\times10^{-7}$). 
This GW peak signal is placed within the sensitivity range of space-based interferometers such as LISA (See the right panel of Fig.~\ref{walking-figure}). 
\begin{table}[htbp!]
	\caption{\label{tab:B-L}%
		Predicted quantities, the same ones as those listed in Table~\ref{tab:walking}, for the $U(1)_{B-L}$ gauge extension of the SM with the CSI. 	}
	\begin{ruledtabular} 
		\begin{tabular}{lc|cccccc}
			 & $m_0/v_\phi$ & $T_n[{\rm MeV}]$ & $T_p[{\rm MeV}]$ &
			$\alpha$ &
			$\beta_H$ 
            &
			$M_{\rm PBH}$ &
			$f_{\rm PBH}$\\
			\colrule
			\text{BP B} & 0 & $145.12$ & $104.37$ &  $1.8\times 10^{6}$ & $7.53$ 
            & $2.041\times 10^{-7}M_\odot$ & $ 0.29$ \\
			BP B$'$ & $5.4\times 10^{-7}$ & $145.12$ & $104.37$ & $1.8\times 10^{6}$ & $7.53$ 
            & $2.041\times 10^{-7}M_\odot$ & $1.03\times10^{-5}$
		\end{tabular}
	\end{ruledtabular}
\end{table}

\begin{figure}[htbp]
	\centering
	\includegraphics[scale=0.55]{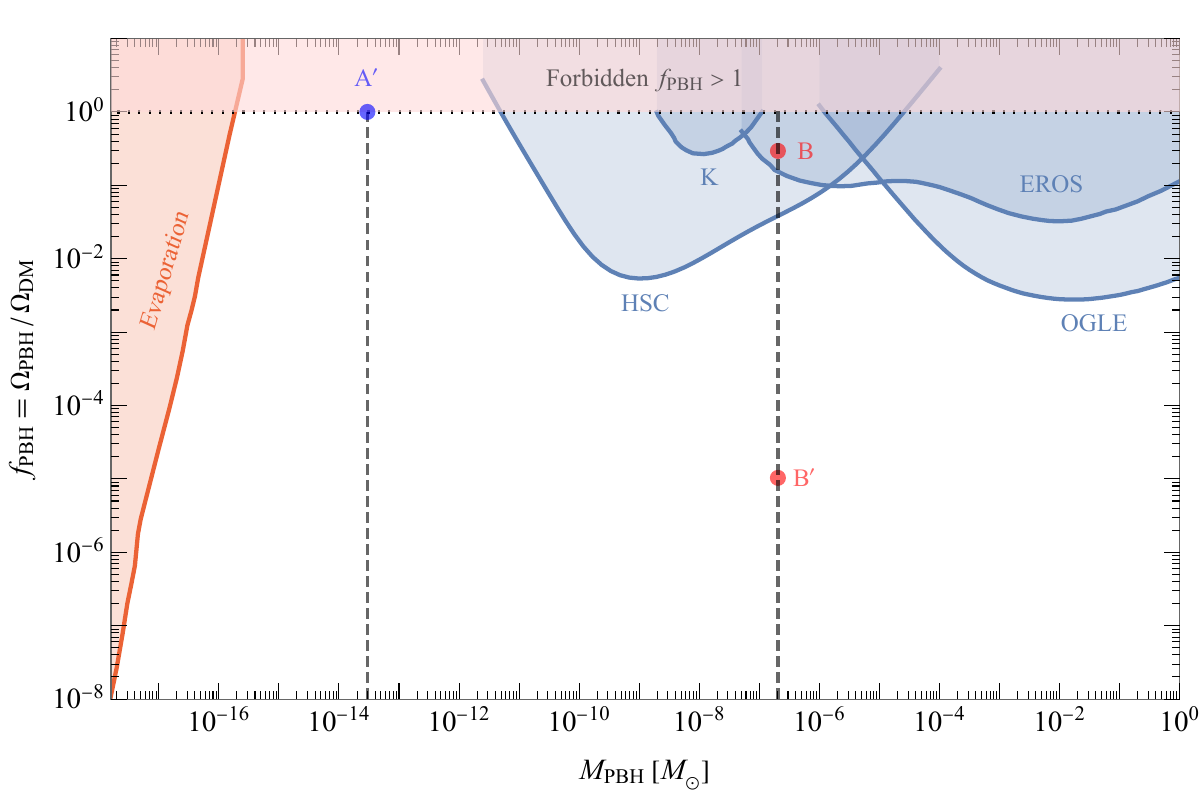}
	\caption{PBH abundances as a function of mass for the benchmark points A$'$, B, and B$'$ given in Tabs.~\ref{tab:walking} and \ref{tab:B-L}. The red shaded regions indicate current constraints from Hawking evaporation, including EGRB~\cite{Carr:2009jm}, 511 keV~\cite{Laha:2019ssq}, CMB~\cite{Clark:2016nst}, EDGES 21cm~\cite{Mittal:2021egv}, Voyager I~\cite{Boudaud:2018hqb}, INTEGRAL~\cite{Laha:2020ivk}, Dwarf heating~\cite{Kim:2020ngi}, Super K~\cite{Dasgupta:2019cae}, COMPTEL~\cite{Coogan:2020tuf}, AMS-02~\cite{Su:2024hrp}, X-ray BG~\cite{Tan:2024nbx}, and Lyman-$\alpha$~\cite{Khan:2025kag}.
    The blue shaded regions correspond to existing microlensing limits from Hyper Suprime-Cam (HSC)~\cite{Niikura:2017zjd,Smyth:2019whb,Croon:2020ouk}, Kepler (K)~\cite{Griest:2013esa,Griest:2013aaa}, EROS~\cite{EROS-2:2006ryy}, and OGLE~\cite{Niikura:2019kqi,Mroz:2024mse,Mroz:2024wag,Mroz:2017mvf}. 
    This figure has been created with the publicly available \texttt{Python} code \texttt{PBHbounds}~\cite{Kavanagh:2024lgq}.}
	\label{PBH-constraints}
	
\end{figure}

In Fig.~\ref{PBH-constraints}, we summarize the predicted PBH abundances  
for two benchmark points BP A$'$ (see Table~\ref{tab:walking}) and BP B$'$ 
(see Table~\ref{tab:B-L}), in comparison with the case in absence of the soft-scale 
breaking mass (``BP B" case with still sizable $f_{\rm PBH}$ yielded). 
Thus, in contrast to the scenario with a positive soft-scale breaking term, 
the negative soft-scale breaking case suppresses the PBH abundance by an order of magnitude, as depicted in Fig.~\ref{PBH-constraints}. 
This destructive contribution accelerates the phase transition, hence enlarges 
the viable parameter space, particularly in the region corresponding to planet-mass PBHs constrained by current microlensing observations. 

\section{Summary and discussion}\label{sec:summary}

In summary, 
\if{we have discussed a possible scenario for the creation of PBHs in a classical scale invariant model. 
The physical scales are generated both dynamically due to hidden strong dynamics and spontaneously due to the CW potential. 
Thanks to the flatness of the CW potential, 
the time of the dilaton at around the false vacuum becomes longer than in the conventional double-well type potential, 
then the PBHs are formed. 
Since the universe is dominated by the vacuum energy of the false vacuum, 
the supercooling occurs during the formation of PBHs. 
In order to produce an appropriate amount of PBHs to be consistent with the observed amount of the dark matter, 
it is necessary to add a mass term that explicitly breaks scale invariance. 
The impact of such a mass term on the PBH abundance was the subject of a quantitative analysis, after which the dynamical origin of the mass term was addressed. 
Consequently, the observed amount of the dark matter can be fully explained by PBHs depending on the choice of the size of the additional mass parameter. 
}\fi 
we have discussed the PBH production arising from the supercooling phase transition for 
the scale symmetry breaking based on the effective potential of CW type. 
We have investigated the effect on the PBH production in the presence of 
a soft-scale breaking mass term for the CW-scalar field, 
which serves as the extra explicit-scale breaking term other than the quantum 
scale anomaly induced by the CW mechanism. 
It turned out that even a small size of the soft-scale breaking term  
can significantly affect the PBH production depending on its sign.  
A positive mass-squared deformation modifies the shape of the bounce action curve from a tangent-like profile to a skewed U-shape, reflecting a delayed tunneling process. This prolongs the duration of the phase transition (i.e., reduces $\beta_H$) and consequently suppresses PBH formation. In contrast, a negative deformation sharpens the bounce profile, making it steeper and accelerating the phase transition, which in turn enhances PBH production. As illustrated in Tabs.~\ref{tab:table2} and \ref{tab:table3}, even tiny variations in the breaking parameter—at the level of $\mathcal{O}(10^{-6})$—can shift the predicted PBH abundance by many orders of magnitude. Such changes can elevate the abundance from negligible to a level sufficient to account for the entirety of dark matter, or reduce it from a severely overproduced regime into the observationally viable window. This mechanism broadens the parameter space and improves model flexibility in explaining dark matter.

We substantiate our results with two concrete scenarios: i) a many-flavor QCD scenario 
(Sec.~\ref{sec:walking-model}); 
ii) a $U(1)_{B-L}$ extension of the SM with CSI (Sec.~\ref{sec:B-L-model}). 
These examples make it manifest that both positive and negative soft-scale breaking mass terms can be naturally realized within well-motivated particle physics models, offering concrete mechanisms behind the parameter deformations that control PBH production.


The three-dimensional effective theory analysis~\cite{Croon:2020cgk,Kierkla:2023von,Kierkla:2025vwp,Kierkla:2025qyz} may yield quantitatively different results. A thorough reanalysis incorporating these effects, as well as recent refinements of PBH abundance estimates accounting for trapped false vacuum domains, nucleation time fluctuations of multiple bubbles, horizon-scale fluctuations during thermal inflation, and higher-order corrections to the bubble nucleation rate~\cite{Lewicki:2023ioy,Lewicki:2024ghw,Lewicki:2024sfw}, lies beyond the scope of this work and will be pursued in future studies. Here, we focus on the impact of soft-scale breaking on PBH formation using a simpler approach, which we expect to capture the essential physics.

In addition to the PBH formation, the GWs can also be produced by the ultra-supercooling. The predicted spectra for representative benchmark points fall squarely within the projected sensitivity of upcoming space‑based interferometers such as LISA, DECIGO, and BBO, providing an independent and testable signature of the underlying CSI dynamics 
(See also Fig.~\ref{walking-figure}). 
Our present study would thus also provide a new testable link between PBH dark matter and GW  signatures in the CW-type scenario.

\section*{Acknowledgments} 

We thank Mei Huang and Ke-Pan Xie for useful suggestions and discussions. 
This work was supported in part by 
the Seeds Funding of Jilin University (S.M.), 
JSPS KAKENHI Grant Number 24K07023 (H.I.), Grant-in-Aid for Research Activity Start-up 23K19052 (K.H.), and Grant-in-Aid for Early-Career Scientists 25K17398 (K.H.).

\appendix 

\section{Derivation of thermal effective potential Eq.~(\ref{eq:veffT})} 
\label{LSM}

In this appendix, following the procedure in the literature~\cite{Miura:2018dsy,Zhang:2023acu,Zhang:2024vpp} 
we briefly describe the steps in the derivation of Eq.~(\ref{eq:veffT}) based on a linear sigma model having $U(N_f)_L\times U(N_f)_R$ chiral symmetry.

The linear sigma model 
with the classical scale invariance takes the form
\begin{equation}
    \mathcal{L} = \text{Tr}\left(\partial_\mu M^\dagger\partial^\mu M\right) - \lambda_1\left[\text{Tr}\left(M^\dagger M\right)\right]^2 - \lambda_2 \text{Tr}\left[\left(M^\dagger M\right)^2\right]\,,
\end{equation}
where the $N_f\times N_f$ matrix field $M_{ab}$ can be decomposed into $N_f^2$ scalar mesons $s^a$ and $N_f^2$ pseudoscalar mesons $p^a$:
\begin{equation}
	M=\sum_{a=0}^{N_{f}^{2}-1}\left(s^{a}+i p^{a}\right) T^{a}\,,
	\label{eq:matrixfield}
\end{equation}
with $T^0=\frac{{\bf 1}}{\sqrt{2N_f}} \mathbb{I}_{_{N_f\times N_f}}$ and $T^i$ being generators of $SU(N_f)\, (i=1, \cdots,N_f^2-1)$ normalized as ${\rm Tr} [T^i T^j] = 
\delta^{ij}/2 $. 
The Lagrangian is invariant 
under $U(N_f)_L \times U(N_f)_R$ chiral transformation for $M$ as 
\begin{align}
	M \to  g_L \cdot M \cdot g_R^\dagger, \quad g_L, g_R \in U(N_f)
\,. 
\end{align} 
The $M$ is assumed to develop the vacuum expectation value (VEV) along the $U(N_f)$ singlet direction, i.e., $s^0$, which reflects the underlying 
large $N_f$ QCD nature as the vector-like gauge theory.

Through the analysis of the renormalization group (RG) equations, 
the GW mechanism~\cite{Gildener:1976ih} tells us that if one takes the condition $\lambda_1=-\lambda_2/N_f$ at some RG scale $\mu_{\rm GW}$~\cite{Miura:2018dsy,Kikukawa:2007zk}, 
there exists a flat direction in the tree-level potential identically vanishes and a massless scalar emerges (dubbed the scalon), along which perturbation theory can be used. 
Thus, the radiative corrections along the flat direction develop a nontrivial vacuum away from the origin, 
a false vacuum as the consequence of the scale anomaly associated with the introduced RG scale. 
With a suitable renormalization condition, the one-loop potential $V_1$ 
in the present linear sigma model can thus be calculated as~\cite{Miura:2018dsy,Kikukawa:2007zk} 
\begin{widetext}
	\begin{align}
		V_1(M) = \frac{1}{64\pi^2} \sum_{a=0}^{N_f^2-1} 
		\left(
		m_{s^a}^4(M) \left( \ln{\frac{m_{s^a}^2(M)}{\mu_{_{\rm GW}}^2}} - \frac{3}{2} \right)
		+
		m_{p^a}^4(M) \left( \ln{\frac{m_{p^a}^2(M)}{\mu_{_{\rm GW}}^2}} - \frac{3}{2} \right)
		\right)\,,
	\end{align}
\end{widetext}
where $m_{s^a}^2$ and $m_{p^a}^2$ are the mass functions 
for scalars and pseudoscalars:  
\begin{align}
	m_{s^a}^2 = \frac{\partial^2 V_0(M)}{\partial (s^a)^2}\,, 
	\quad \quad
	m_{p^a}^2 =\frac{\partial^2 V_0(M)}{\partial (p^a)^2}\,. 
\end{align}  
By means of the chiral rotation, it is possible to choose $s^0$ to be the flat direction as 
\begin{align}
	\langle M 
	\rangle = 
	T^0 \langle s^0\rangle =\frac{1}{\sqrt{2 N_f} } \mathbb{I} \cdot  \langle s^0\rangle\,.
\end{align} 
 Then $m_{s^a}^2$ and $m_{p^a}^2$ 
 can be expressed as
\begin{align}
	\label{eq:mass}
	m_{s^0}^2(s^0) &= 0\,, 
	& 
	m_{s^i}^2(s^0) &= \left(\lambda_1 +\lambda_2\frac{3}{N_f}\right)(s^0)^2=\frac{2 \lambda_2}{N_f} (s^0)^2\,, 
	\notag \\
	m_{p^0}^2(s^0) &= 0\,, 
	& 
	m_{p^i}^2(s^0) &=0\,, 
\end{align} 
where the flat direction condition $\lambda_1 + \lambda_2/N_f =0$ has been used. 
There are two types of the Nambu Goldstone (NG) bosons: one is the scalon, $s^0$, associated with the spontaneous breaking of the scale symmetry along the flat direction; the other consists of the NG bosons, $p^a$, arising from the spontaneous chiral symmetry breaking $\text{U}(N_f)_L\times \text{U}(N_f)_R \rightarrow \text{U}(N_f)_V$. 
Therefore, the effective potential for the scalon $s^0$ is given by
\begin{equation}
V_{\rm eff}(s^0) = 
\frac{N_f^2-1}{64\pi^2} m_{s^i}^4(s^0)
\left( \ln{\frac{m_{s^i}^2(s^0)}{\mu_{_{\rm GW}}^2}} - \frac{3}{2} \right)\,.
\label{eq:veff-0}
\end{equation}

At the finite temperature, along the flat direction, 
the thermal corrections to the walking dilaton potential 
can be evaluated by computing the thermal loops in which only 
the heavy scalar mesons $s^i$ flow. 
Taking into account also the higher loop corrections via the so-called daisy resummation, 
we thus get 
\begin{equation}
V_{\rm eff}(s^0,T) 
= \frac{N_f^2-1}{64\pi^2} 
\mathcal{M}_{s^i}^4(s^0, T)
\left( \ln{\frac{
		\mathcal{M}_{s^i}^2(s^0, T)
	}{\mu_{_{GW}}^2}} - \frac{3}{2} \right) 
+\frac{T^4}{2\pi^2} (N_f^2-1) J_B\left(
\mathcal{M}_{s^i}^2(s^0, T)
/T^2\right)\,,
\end{equation} 
where the $s^i$ scalar meson masses have been dressed as 
$\mathcal{M}_{s^i}^2(s^0, T) = m_{s^i}^2(s^0) + \Pi(T)$ with
\begin{align}
\Pi(T) = \frac{T^2}{6}\bigl((N_f^2 + 1)\lambda_1 + 2N_f \lambda_2\bigr)\Big|_{\lambda_1 = -\lambda_2 / N_f} \,.
\end{align}

\section{Gravitational waves}\label{sec:GWs}

In this Appendix, we discuss the generation of stochastic gravitational wave backgrounds sourced from the ultra-supercooling. 
The resulting GW spectrum, ($\Omega_{\rm GW}h^2$), arises from three processes: bubble wall collisions ($\Omega_{\rm coll}h^2$), sound waves in the plasma ($\Omega_{\rm sw}h^2$), and magnetohydrodynamic turbulence in the plasma ($\Omega_{\rm turb}h^2$), i.e.,
$$\Omega_{\rm GW}h^2 \simeq \Omega_{\rm coll}h^2 + \Omega_{\rm sw}h^2 +\Omega_{\rm turb}h^2\,.$$
If bubble collisions occur before the walls reach their terminal velocity, most of the vacuum energy is consumed to accelerate the bubble walls, and the GW spectrum is dominated by the scalar field contribution. In contrast, if the bubbles reach their terminal velocity, most of the vacuum energy is transferred to the surrounding plasma rather than the bubble walls, and sound waves dominate over the spectrum.
The contribution from turbulence is expected to be subdominant, we neglect it in what follows.

The efficiency factor for bubble collisions $\kappa_{\text{coll}}\equiv E_{\rm wall}/E_{\rm V}$, which represents the fraction of the vacuum energy that goes into the bubble walls, is given by~\cite{Azatov:2019png}
\begin{equation}
    \kappa_{\text{coll}} = 
\begin{cases} 
\displaystyle \frac{\gamma_{\text{eq}}}{\gamma_*} \left[ 1 - \frac{\alpha_{\infty}}{\alpha} \left( \frac{\gamma_{\text{eq}}}{\gamma_*} \right)^2 \right], & \gamma_* > \gamma_{\text{eq}} \\
1 - \frac{\alpha_{\infty}}{\alpha}, & \gamma_* \leq \gamma_{\text{eq}}\,,
\end{cases}
\end{equation}
The Lorentz factors for the bubble velocity at equilibrium, $\gamma_{\text{eq}}$~\cite{Bodeker:2009qy,Bodeker:2017cim,Gouttenoire:2021kjv}, and at collision, $\gamma_\star$~\cite{Ellis:2019oqb}, are defined as follows:
\begin{equation}
    \gamma_{\text{eq}} = \frac{\alpha - \alpha_{\infty}}{\alpha_{\text{eq}}}, \quad \text{and} \quad \gamma_\star= \frac{2}{3}\frac{R_\star}{R_0},
\end{equation}
where $R_0$ is the initial bubble radius~\cite{Ellis:2019oqb}
\begin{equation}
    R_0 = \left[\frac{3S_3(T_n)}{4\pi \Delta V}\right]^{1/3}\,,
\end{equation}
and 
$R_\star$ is the radius at collision, which is derived from the bubble number density $n_B$ as 
\begin{equation}
    R_\star = n_B(T_\star)^{-1/3} =\left[ T_\star^3 \int_{T_\star}^{T_c} \frac{dT}{T^4} \frac{\Gamma(T)}{H(T)} e^{-I(T)} \right]^{-1/3},
\end{equation}
with
\begin{equation}
    I(T) = \frac{4\pi}{3} \int_{T}^{T_c} \frac{\mathrm{d}T' \, \Gamma(T')}{T'^4 H(T')} \left( \int_{T}^{T'} \frac{v_w\,\mathrm{d}T''}{H(T'')} \right)^3.
\end{equation}
In the exponential nucleation approximation, $\beta$ is a good estimate of both the inverse time duration of the phase transition and the characteristic frequency of the GWs, through the mean bubble separation. In that case, $R_\star$ is evaluated as 
\begin{equation}
    R_\star = (8\pi)^{1/3}\frac{v_w}{\beta}\,,
\end{equation} 
where $v_w$ is the bubble wall velocity. 

The quantities $\alpha_\infty$ and $\alpha_{\rm eq}$ are associated with the leading-order (LO) and next-to-leading-order (NLO) pressures~\cite{Ellis:2019oqb}
\begin{equation}
     \alpha_{\infty} \equiv  \frac{\Delta P_{\rm LO}}{\rho_R} = \frac{1}{24} \frac{\Delta m^2 T_p^2}{\rho_R} \,, \quad \text{and} \quad \alpha_{\text{eq}} \equiv \frac{\Delta P_{\rm NLO}}{\rho_R} =\frac{g^2 \Delta m_V T_p^3}{\rho_R}\,,
\end{equation}
with 
\begin{align} 
\Delta m^2 &\equiv \sum_i c_i N_i (m_{i,t}^2 - m_{i,f}^2) 
\,, 
\notag \\ 
g^2 \Delta m_V &\equiv \sum_{i\in V} g_i^2 N_i (m_{i,t} - m_{i,f})
\,. 
\end{align}  
Here, $N_i$ is the number of degrees of freedom of particle $i$, $c_i=1\,(1/2)$ for bosons (fermions). The sum runs over all particle species $i$ that are light in the false vacuum and become heavy in the true vacuum, and in the latter case only gauge bosons are included, where $g_i$ are their respective gauge couplings.

The GW spectrum from bubble wall collisions is captured by the
 envelope approximation with the fitting formula~\cite{Caprini:2015zlo,Huber:2008hg}
\begin{equation}
    \Omega_{\text{coll}}h^2= 1.67 \times 10^{-5} \left( \frac{H_\star R_\star}{(8\pi)^{1/3}} \right)^2 \left( \frac{\kappa_{\rm coll} \alpha}{1 + \alpha} \right)^2 \left( \frac{100}{g_\star} \right)^{1/3}  \frac{0.11 v_w}{0.42 + v_w^2} \frac{3.8 (f / f_{\text{coll}})^{2.8}}{1 + 2.8 (f / f_{\text{coll}})^{3.8}} \, ,
\end{equation}
where the peak frequency is given by
\begin{equation}
    f_{\text{coll}} = 1.65 \times 10^{-5} \text{Hz} \frac{(8\pi)^{1/3}}{H_\star R_\star}\left( \frac{0.62v_w}{1.8 - 0.1 v_w + v_w^2} \right) \left( \frac{T_{\rm reh}}{100 \text{ GeV}} \right) \left( \frac{g_\star}{100} \right)^{1/6}\,.
\end{equation}

The GW spectrum from fluid motions is dominated by sound waves fitted by numerical simulations~\cite{Hindmarsh:2013xza,Hindmarsh:2015qta,Hindmarsh:2017gnf} as
\begin{equation}
    \Omega_{\text{sw}}h^2 = 1.64\times10^{-6} (H_\star \tau_{\text{sw}})(H_\star R_\star)\left( \frac{\kappa_{\text{sw}} \alpha}{1 + \alpha} \right)^2 ( f/f_{\text{sw}})^3 \left[ \frac{7}{4+3(f/f_{\rm sw})^2} \right]^{7/2}\,,
\end{equation}
with the peak frequency
\begin{equation}
    f_{\text{sw}} = 2.6\times 10^{-5}\text{Hz}\frac{1}{H_\star R_\star}\left(\frac{T_{\rm reh}}{100{\rm GeV}}\right)\left(\frac{g_\star}{100}\right)^{1/6}\,,
\end{equation}
and the lifetime of sound waves in units of Hubble time is $H_\star\tau_{\rm sw}={\rm min}\left(1,2H_\star R_\star/\sqrt{3K}\right)$, where $K=0.6\kappa_{\rm sw}\alpha/(1+\alpha)$.
The efficiency factor $\kappa_{\text{sw}}$ which characterizes the amount of released vacuum energy into the kinetic energy of fluid motions is given by~\cite{Ellis:2020nnr}
\begin{equation}
    \kappa_{\text{sw}} = \frac{\alpha_{\text{eff}}}{\alpha} \frac{\alpha_{\text{eff}}}{0.73 + 0.083 \sqrt{\alpha_{\text{eff}}} + \alpha_{\text{eff}}}\,,
\end{equation}
with $\alpha_{\rm eff}=\alpha(1-\kappa_{\rm coll})$.

When the resulting PBHs lie within the asteroid-mass range (\ref{asteroid-mass}) of particular interest, and the stochastic GW background is dominated by sound waves generated during the PT, the corresponding peak frequency $f_{\rm sw}$ is expected to fall in the range $\sim 1\text{ mHz}-10\text{ Hz}$, while the peak amplitude $\Omega_{\rm sw}^{\rm peak} h^2$ reaches values around $\sim 10^{-6}$. This places the GW signal well within the sensitivity range of upcoming space-based interferometers such as LISA, DECIGO, and BBO. Such a signal offers a promising target for future GW observations, providing a potential probe of the underlying PT dynamics and PBH formation mechanism.


\end{document}